\documentclass[12pt]{article}

\usepackage{theorem,amssymb,amsbsy,latexsym}

\theoremstyle{plain}

\newtheorem{Lem}{Lemma}
\newtheorem{Prop}{Proposition}
{\theoremheaderfont{\it}\theorembodyfont{\upshape}\newtheorem{Rem}{Remark}[section]}

\newcommand{\al}{\underline{\alpha}}
\newcommand{\be}{\underline{\beta}}
\newcommand{\tet}{\theta}

\newcommand{\Tr}{{\rm Tr}}
\newcommand{\tW}{\tilde{W}}

\newcommand{\rr}{{\cal R}}
\newcommand{\ff}[2]{\mbox{$\f{{#1}}{{#2}}$}}
\newcommand{\teps}{\tilde\eps}
\newcommand{\pr}[2]{\left< {#1},{#2}\right>}
\newcommand{\ga}{\gamma} 
\newcommand{\id}{{\bf 1}}

\newcommand{\jj}{j} 
\newcommand{\hj}{\hat{j}} 

\newcommand{\ccr}[2]{\left[{#1},{#2}\right]}

\newcommand{\ee}[1]{{\rm e}^{#1}}
\newcommand{\ii}{{\rm i}}
\newcommand{\f}{\frac}

\newcommand{\vx}{{\bf x}}
\newcommand{\vy}{{\bf y}}

\newcommand{\vn}{{\bf n}}

\newcommand{\vE}{{\bf E}}
\newcommand{\vP}{{\bf P}}

\newcommand{\Ref}[1]{(\ref{#1})}

\renewcommand{\b}{{\hat \rho}}
\newcommand{\bb}{\tilde{\rho}}
\newcommand{\U}{{\rm U}}
\newcommand{\sgn}{{\rm sgn}}
\newcommand{\eps}{\varepsilon}

\newcommand{\xxa}{\stackrel {\scriptscriptstyle \times}{\scriptscriptstyle \times} \!}
\newcommand{\xxe}{\! \stackrel {\scriptscriptstyle \times}{\scriptscriptstyle \times}}

\newcommand{\half}{\mbox{$\frac{1}{2}$}}

\newcommand{\R}{{\mathbb R}}
\newcommand{\C}{{\mathbb C}}
\newcommand{\Z}{{\mathbb Z}}
\newcommand{\Zp}{\Z\backslash\{ 0\}}
\newcommand{\N}{{\mathbb N}}

\newcommand{\cA}{{\cal A}}
\newcommand{\cC}{{\cal C}}
\newcommand{\cD}{{\cal D}}

\newcommand{\cP}{{\cal P}}
\newcommand{\cH}{{\cal H}}

\newcommand{\cF}{{\cal F}}
\newcommand{\cE}{{\cal E}}
\newcommand{\cG}{{\cal G}}

\newcommand{\cX}{{\cal X}}

\newcommand{\cZ}{{\cal Z}}

\newcommand{\QED}{\hfill$\square$}

\newcommand{\eq}{\begin{equation}}
\newcommand{\eqend}{\end{equation}}
\newcommand{\eqa}{\begin{eqnarray}}
\newcommand{\nonueqa}{\begin{eqnarray*}}
\newcommand{\eqaend}{\end{eqnarray}}
\newcommand{\nonueqaend}{\end{eqnarray*}}
\newcommand{\nonu}{\nonumber \\ \nopagebreak}
\newcommand{\bma}[1]{\begin{array}{#1}}
\newcommand{\ema}{\end{array}}
\newcommand{\bc}{\begin{center}}
\newcommand{\ec}{\end{center}}

\begin{document}
\begin{flushright}
February 7, 2001
\end{flushright}
\vspace{.4cm}

\begin{center}

{\Large \bf Second quantization of the elliptic
Calogero-Sutherland model}\\
\vspace{1 cm}

{\large Edwin Langmann}\\
\vspace{0.3 cm}
{\em Theoretical Physics, Royal Institute of Technology, S-10044 Sweden}\\
\end{center}

\begin{abstract}
We construct a quantum field theory model of anyons on a circle and at
finite temperature. We find an anyon Hamiltonian providing a second
quantization of the elliptic Calogero-Sutherland model. This allows us
to prove a remarkable identity which is a starting point for an
algorithm to construct eigenfunctions and eigenvalues of the elliptic
Calogero-Sutherland Hamiltonian.

\bigskip

\noindent {PACS: 71.10.Pm, 05.30.Pr, 02.20.Tw \newline
 MSC-class: 81T10, 81V70, 17B69}
\end{abstract}

\section{Introduction}
This is the first of two papers on the elliptic Calogero-Sutherland
model, providing details and proofs of various results announced in
Ref.\ \cite{EL0} and culminating in an algorithm to solve this model
\cite{EL2}.  The results here are based on loop group techniques: we
relate this model to a quantum field theory model of anyons which, as
we believe, is interesting also in its own right.  We have attempted
to keep our discussion reasonably self-contained (and in particular
explain all the physics terminology we use) and make it accessible
also for mathematicians.

In this Chapter we specify what we mean by finite temperature anyons
and second quantization, and we also provide some background to the
elliptic Calogero-Sutherland model.  We then give the plan for the
rest of the paper by summarizing our results.

\noindent {\bf Background:} 
By {\em anyons} we mean quantum fields $\phi(x)$ parametrized by a
coordinate $x$ on the circle, $-\pi\leq x < \pi$, obeying
exchange relations
\eq 
\label{a1}
\phi(x)\phi(y) =
\ee{\pm \ii\pi\lambda}\phi(y)\phi(x) 
\eqend 
where $\lambda>0$ is the so-called {\em statistics parameter}.  These
anyons generalize bosons and fermions which correspond to the special
cases where the phase factor $\ee{\pm \ii\pi\lambda}$ is $+1$ and
$-1$, respectively.\footnote{The name {\em any}ons is due to the fact
that this can be {\em any} phase in general.} Mathematically, these
anyons are operator valued distributions on some Hilbert space $\cF$,
and our construction amounts to giving a precise mathematical meaning
to these objects and defining and computing {\em anyon correlation
functions}. The latter are given by a linear functional
$\omega$ on the $*$-algebra generated by the anyons (the $*$ is
the Hilbert space adjoint). We construct a particular
representation of the anyons which is such that the anyon correlation
functions are given by elliptic functions with the {\em nome}
\eq
\label{q}
q=\exp{(-\beta/2)}\, , \quad \beta >0 
\eqend 
where $\beta$ can be interpreted as {\em inverse
temperature}. For example, the simplest non-trivial anyon correlation
function which we obtain is
\eq
\label{OmOm}
\omega( \phi(x)^* \phi(y) )  = const.\; \tet(x-y)^{-\lambda} 
\eqend
where 
\eq
\label{tet}
\tet(r) = \sin(r/2) \prod_{n=1}^\infty (1-2q^{2n}\cos(r) + q^{4n})
\eqend 
is equal, up to a multiplicative constant, to the Jacoby Theta
function $\vartheta_1(r/2)$.  This function, or rather its regularized
version defined in Eq.\ \Ref{b1}, will play a prominent role in this
paper.  We also mention a particular element $\Omega$ in the anyon
Hilbert space $\cF$ which can be interpreted as the {\em thermal
vacuum} since it allows to compute the anyon correlation functions as
{\em vacuum expectation value}, i.e., $\omega(\cdot) = \; \pr{\Omega}{
\cdot \; \Omega}$ ($\pr{\cdot}{\cdot}$ is the inner product in the
anyon Hilbert space). Our construction provides another example of a
quantum field theory model which can be made mathematically precise
using the representation theory of loop groups (see e.g.\
\cite{PS}). A main difficulty is what physicists call {\it
ultra-violet divergences}: the anyons $\phi^{(*)}(x)$ are operator
valued distributions and thus products of them need to be defined with
care. For example, the relation in Eq.\ \Ref{a1} becomes problematic
for $x=y$, and this difficulty also manifests itself in the anyon
correlation function in Eq.\ \Ref{OmOm} which is singular for $x\to
y$. The approach we use provides a particularly simple solution to
this problem (we will describe the main idea in the Summary below).

It is interesting to note that this very same quantum field theory
model of anyons, for odd integers $\lambda$, has been used in the
theory of the fractional quantum Hall effect \cite{Wen}. We also note
that different constructions of finite temperature anyons on the real
line were recently given in \cite{IT,LMP}, and (many of) the results
there can be (formally) obtained from ours by rescaling variables
$x\to 2\pi x/L$ and $\beta\to 2\pi \beta/L$ (which changes the
circumference of the circle from $2\pi$ to an arbitrary length $L>0$)
and taking the limit $L\to \infty$.

We now give some background to the elliptic Calogero-Sutherland (eCS)
model.  The eCS model is defined by the differential
operator
\eq
\label{eCS}
H_{N} =  - \sum_{j=1}^N\frac{\partial^2}{\partial x_j^2} 
\; + \;   
\ga\!\!  \sum_{1\leq j<k\leq N}  V(x_j-x_k)
\eqend 
with $-\pi\leq x_j \leq\pi$ coordinates on the circle $S^1$,
$N=2,3,\ldots$, $\ga >-1/2$, and 
\eq
\label{V}
V(r) \, = \, -\f{\partial^2}{\partial r^2} \log \tet(r) 
\, , 
\eqend 
$\tet$ as in Eq.\ \Ref{tet}. This function $V(r)$ is equal, up to an
additive constant, to Weierstrass' elliptic function $\wp(r)$ with
periods $2\pi$ and $\ii \beta$ (see Eq.\ \Ref{wp} in Appendix~A for
the precise formula). This differential operator defines a selfadjoint
operator on the Hilbert space of square integrable functions on
$[-\pi,\pi]^N$ which provides a quantum mechanical model of $N$
identical particles moving on a circle of length $2\pi$ and
interacting with a two body potential proportional to $V(r)$ where
$\ga$ is the coupling constant.\footnote{As will be discussed in more
detail in \cite{EL2}, this model corresponds to a particular
self-adjoint extension \cite{KT} which, for positive $\ga$, is the Friedrich's
extension of $H_N$; see e.g.\ Theorem X.23 in Ref.\ \cite{RS2}.}  This
model is a prominent integrable many body system (a standard review is
Ref.\ \cite{OP}). In particular the limiting case $q=0$ where the
interaction potential becomes a trigonometric functions, $V(r) =
(1/4)\sin^{-2}(r/2)$, is the celebrated Sutherland model whose
complete solution was found about 30 years ago \cite{Su}.  This
explicit solution plays a central role in remarkably many different
topics in theoretical physics including matrix models, quantum chaos,
QCD, and two dimensional quantum gravity (for review see, e.g.,
Ref.\ \cite{Guhr}, Sect.\ 7). There is also an interesting relation
between the Sutherland model and the theory of the fractional quantum
Hall effect (see e.g.\ \cite{IR,Wen,YZZ}) which will be discussed in
more detail below.  We note that eigenfunctions of the eCS
differential operator in Eq.\ \Ref{eCS} are known only for $N=2$
and/or integer values of the coupling parameter\footnote{If we write
$\ga$ as in Eq.\ \Ref{ga} below then the coupling parameter is equal
to $\lambda$.} $(1 + \sqrt{1+2\ga})/2$: For $N=2$ these are classical
results on Lam\'{e}'s equation (see e.g.\ \cite{WW}) which recently
were generalized \cite{EK,R} and extended to $N>2$ \cite{FV1,FV2}.

Of course, the differential operator in Eq.\ \Ref{eCS} does not define
a unique self-adjoint operator, but our approach will automatically
specify a particular self-adjoint extension \cite{EL2} (which for
$q=0$ is identical with the one solved by Sutherland \cite{Su}; we
note that some of the known eigenfunctions of the eCS differential
operator mentioned are singular and do not correspond to that
particular self-adjoint extension).

By a {\em second quantization} of the eCS model we mean one operator
$\cH=\cH^*$ on the anyon Hilbert space $\cF$ which accounts for the
eCS Hamiltonians $H_N$ in Eq.\ \Ref{eCS} for all particle numbers
$N$. To be more specific, the commutator of this operator $\cH$ with a
product of $N$ anyons
\eq \Phi^N(\vx) \, := \, \phi(x_1)
\cdots \phi(x_N) 
\eqend 
is essentially equal to the eCS Hamiltonian
applied to this very product, i.e.,
\eq
\label{sq1_I}
\ccr{\cH}{\Phi^N(\vx)}\Omega =  H_N \Phi^N (\vx)\Omega 
\eqend
where $\Omega$ is the above-mentioned thermal vacuum, and the coupling
constant of the eCS model is determined by the statistics parameter of
the anyons as follows,
\eq
\label{ga}
\ga = 2\lambda(\lambda-1)\, .  \eqend 
Such a second quantization was previously known in the trigonometric
limit (corresponding to zero temperature) \cite{MP,AMOS,I,MS,CL},
and in this paper we generalize it to the elliptic case.  It is
remarkable that this generalization is most natural from a physical
point of view: it amounts to going from zero- to finite
temperature. In the trigonometric limit (corresponding to zero
temperature), this second quantization has provided an interesting
direct link between the Sutherland model and the theory of the
fractional quantum Hall effect, and we expect that our finite
temperature generalization should be interesting in this context,
too. However, our main motivation and emphasis is mathematical: In
Ref.\ \cite{CL} the second quantization of the Sutherland model was
used to derive an algorithm for constructing eigenvalues and
eigenfunctions of the Sutherland model and thus recover the solution
of Sutherland \cite{Su}.  We will use the second quantization to
derive a remarkable identity for the anyon correlation function
$F_N(\vx,\vy)=\pr{\Omega}{\Phi^N(\vx)^*\Phi(\vy)\Omega}$ (see
Proposition~\ref{P3}) which, as we will outline in the conclusions of
the paper, is the starting point for a novel algorithm for solving the
eCS model \cite{EL2}.  To obtain this identity we will 
need\footnote{$[a,b]:=ab-ba$}
\eq
\label{sq2_I}
\pr{\Omega}{ [\cH, \Phi^N(\vx)^*\Phi^N(\vy)] \Omega} = 0
\eqend
which is the second important property of $\cH$ and in fact the one
which restricts us to interactions $V(r)$ which are Weierstrass
elliptic functions (Eq.\ \Ref{sq1_I} actually holds true in more
generality).

\bigskip

\noindent {\bf Summary of results:} In Section~2 we construct anyons,
i.e., give a precise mathematical meaning to the quantum fields
$\phi(x)$ and compute all anyon correlation functions.  The basic idea
of our construction is to use vertex operators similar to the ones
used in string theory (see e.g.\ \cite{strings}) and which we make
mathematically precise using the representation theory of the loop
group of U(1) (in the spirit of Ref.\ \cite{Seg}). We deviate from a
similar previous construction of zero temperature anyons \cite{CL} in
that we use a somewhat unusual class of reducible representation of
this loop group which, in special cases of particular interest to us,
can be interpreted as finite temperature representations (the precise
statement and proof of this is given in Appendix~B.3).  Technically,
we account for the distributional nature of the quantum fields
$\phi(x)$ by using a {\em regularization} which, roughly speaking, is
a generalization of the idea to represent distributions as limits of
smooth functions (e.g.\ the delta distribution on the circle as limit
$\eps\downarrow 0$ to the smooth function $\delta_\eps(x) \, = \,
1/(2\pi)\sum_{n\in \Z} \exp{( \ii n x -|n|\eps)}$).  In a similar
manner, we will construct {\em regularized anyons} $\phi_\eps(x)$
which for $\eps>0$ can be multiplied without ambiguities and obey Eq.\
\Ref{a1} only in the limit $\eps\downarrow 0$. We perform that latter
limit at a later point where it can be taken without
difficulty.\footnote{One can interpret $1/\eps$ as {\it ultra violet
cut-off}.} For simplicity, we will regard all quantum fields only as
{\em sesquilinear forms} (using results in the literature, e.g.\ from
Refs.\ \cite{CR,GL}, one can prove that many sesquilinear forms which
we construct can be extended to well-defined operators, but since we
actually do not need these results we will only mention them in
passing).  In Section~2.1 representations of the loop group of U(1)
are defined, and we also collect some important technical results
which we will use throughout the paper (Lemma~\ref{L0}). The
definition and main properties of these regularized anyons, and in
particular explicit formulas for all anyon correlation functions, are
given in Section~2.2 (Proposition~\ref{P1}).

In Section~3 the second quantization of the eCS model is constructed.
We will give an explicit formula for $\cH$ and prove that it indeed
obeys Eqs.\ \Ref{sq1_I} and \Ref{sq2_I}, or rather, a generalization
of these relation for regularized anyons, i.e., with the parameters
$\eps$ inserted.  In particular, we will get, instead of the eCS
differential operator $H_N$ in Eq.\ \Ref{eCS}, a regularized operator
$H_N^{2\eps}$ which, roughly speaking, is obtained by replacing the
singular potential $V(r)$ by a potential $V_{2\eps}(r)$ where the
$1/r^2$-singularity of the Weierstrass $\wp$-function is regularized
to $1/(r+\ii\eps)^2$.  We will explicitly give the $\eps$-corrections
to Eq.\ \Ref{sq1_I}.  These results are summarized in Proposition
\ref{P2}. The proof is by explicit, lengthy computations which we
divide in Lemmas and partly defer to Appendix~C. We note that it is
precisely this $\eps$-regularization which determines the self-adjoint
extension of the eCS differential operator for us (however, this will
only become important for us in the second paper \cite{EL2}).

In Section~4 we derive a remarkable identity (Proposition~\ref{P3})
providing the starting point for constructing eigenvalues and
eigenfunctions of the eCS model. In the conclusions (Section~5) we
only state the theorem underlying this algorithm and outline its proof
based on Propositon \ref{P3} (this algorithm will be elaborated in
Ref.\ \cite{EL2}).

Identities about (regularized) elliptic functions which we need are
collected and proven in Appendix~A.  Appendix~B contains a
self-contained discussion of the relation of our quantum field theory
techniques. The proofs of various Lemmas are collected in Appendix~C.

\bigskip

\noindent {\bf Notation:} All Hilbert spaces considered are separable,
and Hilbert space inner products $\pr{\cdot}{\cdot}$ are linear in the
second and anti-linear in the first argument. We denote as $\C$, $\R$,
$\Z$ the complex, real and integer numbers, $\N$ are the positive
integers, and $\N_0=\N \cup \{ 0\}$.  We
denote as $\bar c$ the complex conjugate of $c\in\C$, and
$|c|=\sqrt{c\bar c}$. We identify elements in $\U(1)$ with phases, 
i.e., $c\in\C$ such that $|c|=1$.

\section{Finite temperature anyons on the circle}

\subsection{Reducible representations of the loop group of $\U(1)$}
In this subsection we set the stage for our construction of anyons. 

We consider the $*$-algebra $\cA_0$ with identity $\id$ 
generated by elements $\b(n)$, $n$ integer, and $R$ obeying the following 
relations, 
\eq
\label{12}
\ccr{\b(m)}{\b(n)} = m\delta_{m,-n}\id \, ,
\quad  \ccr{\b(n)}{R} = \delta_{n,0}R 
\eqend
and
\eq
\label{un}
R^*=R^{-1}\: , \quad  \b(n)^*=\b(-n)
\eqend
for all integers $m,n$. 
We will also use the notation $Q=\b(0)$. Note that Eq.\ \Ref{12}
implies     
\eq
\label{QRQ}
\ee{\ii\alpha Q} R^w = \ee{\ii\alpha} R^w\, \ee{\ii\alpha Q} \quad \forall \alpha\in\R,
w\in\Z 
\eqend
which will be useful for us later on.

\begin{Rem}\label{2.1}
{As explained in Appendix~B, the algebra $\cA_0$ 
defines essentially a central extension of the loop group of $\U(1)$.}
\end{Rem}

We now construct a class of representations of $\cA_0$
using a standard highest weight representation of the auxiliary
$*$-algebra $\cA=\cA_0\otimes\cA_0$ with identity $\id$ 
generated by elements $R_A$ and $\b_A(n)$, $A=1,2$ and $n\in\Z$, defined
by the relations
\eqa
\label{ccr}
\ccr{\b_A(m)}{\b_B(n)} = m\delta_{m,-n}\delta_{A,B}\id \, , \quad 
\ccr{\b_A(n)}{R_B} = \delta_{n,0}\delta_{A,B}R_A 
\eqaend
and 
\eq
\label{star}
R_A^* = R_A^{-1}\: , \quad \b_A(n)^* = \b_A(-n) \eqend for all
integers $m,n$ and $A,B=1,2$.  The representation of $\cA$ is on a
Hilbert space $\cF$ with inner product $\pr{\cdot}{\cdot}$ and
completely characterized by the following conditions, \eq
\label{hw}
\b_A(n)\Omega = 0\quad \forall n\geq 0\, , A=1,2\: , 
\eqend
and 
\eq
\label{ROm}
\pr{\Omega}{ R_1^{w_1} R_2^{w_2}\Omega} = \delta_{{w_1},0} \delta_{{w_2},0}
\quad \forall w_{1,2} \in\Z  
\eqend
where $\Omega\in\cF$ is the highest weight vector and  $*$ is the Hilbert space 
adjoint. Indeed, it is easy to check that the rules above imply that the elements 
\eq
\label{eta}
\eta = \prod_{A=1,2}\prod_{n=1}^\infty 
\f{\b_A(-n)^{m_{A,n}}}{\sqrt{m_{A,n}!n^{m_{A,n}}}} \, 
R_1^{w_1} R_2^{w_2}\, \, \Omega\: , \quad 
m_{A,n}\in\N_0 \: , \sum_{A=1,2} \sum_{n=1}^\infty 
m_{A,n}<\infty \: , w_A \in\Z
\eqend
are orthonormal, and the set $\cD$ of all finite linear combinations
of such elements $\eta$ is a pre-Hilbert space carrying a
$*$-representation of the algebra $\cA$.  The Hilbert space $\cF$ is
defined as the norm completion of $\cD$.

We now observe that 
\eqa
\label{bn}
\pi(R)\, := R_1\, \quad
\pi(Q) = \pi(\b(0))\, :=\b_1(0) \nonu 
\pi(\b(n))  \, := c_n\b_1(n) + s_n \b_2(-n) 
\quad \forall n\in\Zp
\eqaend
($s_n,c_n\in\C$) obviously defines a unitary 
representation $\pi$ of the $*$-algebra $\cA_0$ 
provided that 
\eq
\label{cnsn1}
|c_n|^2-|s_n|^2=1
\eqend
and
\eq
\label{cnsn2}
c_{-n}=\overline{c_n}\: , \quad s_{-n}=\overline{s_n} \eqend for all
non-zero integers $n$.  As explained in Remark \ref{2.4} below, it
natural to also require that 
\eq
\label{cnsn3}
\sum_{n\in\Z} |s_n|^4 <\infty \: .  
\eqend
One choice of particular interest for us is 
\eq
\label{expl}
c_n = \left( \f{1}{1-q^{2|n|}}\right)^{1/2} ,\quad 
s_n = \left( \f{q^{2|n|}}{1-q^{2|n|}}\right)^{1/2}\quad \forall 
n\in\Zp 
\eqend
with $|q| < 1$ and $q^2$ real, even though many of our results hold 
true more generally. 

\begin{Rem}\label{2.2}
It is interesting to note that the representation
$\pi$ with $s_n$ and $c_n$ as in Eq.\ \Ref{expl} and $q=\exp{(-\beta/2)}$
is the finite temperature representation of the $*$-algebra 
$\cA_0$ with temperature $1/\beta$ and the Hamiltonian
\eq
\label{Ham}
H=\f{a}{2}Q^2 + \sum_{n=1}^\infty \b(-n)\b(n) 
\eqend
in the limit $a\to \infty$. The interested reader can
find a precise formulation and proof of this statement 
in Appendix~B.3.
\end{Rem}

Since there is no danger of confusion we 
simplify notation and write $R$, $Q$ and $\b(n)$ short for 
$\pi(R)$, $\pi(Q)$ and $\pi(\b(n))$ in the following. 

We now collect some (standard) technical results which we will need.
We define {\it normal ordering} $\xxa\cdot\xxe$ as the linear map on
the algebra generated by monomials $M$ in the $\b_A(n)$ and $R_1$ by
the following inductive rules,
$$
\xxa R_1^w \xxe \, := \, R_1^w \quad \forall w\in\Z
$$
\eqa
\xxa M \b_A(n) \xxe \, = \, \xxa \b_A(n)M\xxe\, := \, \left\{
\bma{ll} \xxa M\xxe \b_A(n)\quad & \mbox{ if $n>0$} \\
\f{1}{2}\left( \xxa M\xxe \b_A(0) +  \b_A(0)\xxa M\xxe\right) 
 \quad & \mbox{ if $n=0$} \\
 \b_A(n) \xxa M\xxe\quad & \mbox{ if $n<0$} \\
\ema \right. \: . 
\eqaend
Note that these rules and Eqs.\ \Ref{ccr}--\Ref{ROm} imply that 
for arbitrary fixed vectors $\eta,\eta'\in\cD$, the expression
$
\pr{\eta}{  \xxa \b_{A_1}(n_1) \cdots\b_{A_k}(n_k)\xxe \eta'}
$
is non-zero only for a {\em finite} number of different 
combinations $n_1,\ldots,n_k\in\Zp$ and 
$A_1,\ldots,A_k\in\{1,2\}$.                 
This implies that for arbitrary complex numbers $v^{A_1,\ldots,A_k}_k
(n_1,\ldots,n_k)$, the expression 
\eq
{\bf V} \, := \, v_0\id + \sum_{k=1}^\infty 
\sum_{A_1,\ldots,A_k=1,2}
\sum_{n_1,\ldots,n_k\in\Zp} 
v^{A_1,\ldots,A_k}_k
(n_1,\ldots,n_k) \xxa \b_{A_1}(n_1) \cdots\b_{A_k}(n_k)\xxe  
\eqend
is a well-defined sesquilinear form on $\cD$, i.e.,  
$\pr{\eta}{{\bf V} \eta'}$ is finite for all $\eta,\eta'\in\cD$
(since it is always a finite sum there is never any problem 
with convergence). Moreover, for all $w\in\Z$ and $a\in\C$ and 
${\bf V}\, =\,  \xxa {\bf V}\xxe$ as above,
\eq
\xxa R^w \ee{a Q} {\bf V}   \xxe\, = \,  \xxa \ee{a Q} R^w  {\bf V}   \xxe\, = \, 
 \ee{a Q/2} R^w \ee{a Q/2} {\bf V} 
\eqend
is a sesquilinear form on $\cD$. 
In particular, these rules imply the following

\begin{Lem}\label{L0}
For arbitrary complex $\alpha_n$,
\eq
\label{PHI}
\Phi \, := \, \xxa R^w \ee{\ii J(\al)}\xxe \quad \mbox{ with } J(\al)
\, :=\, \sum_{n\in\Z}\alpha_n\b(-n)\quad \mbox{ and }  w\in\Z
\eqend 
is a well-defined sesquilinear form on $\cD$ equal to
\eq
\Phi = \ee{\ii\alpha_0 Q/2 }R^w\ee{\ii\alpha_0 Q/2}\ee{\ii J^+(\al)}
\ee{\ii J^-(\al) } \, , 
\quad J^\pm(\al) = \sum_{n=1}^\infty \left[ \alpha_{\pm n} c_{\pm n} \b_1(\mp n)
+ \alpha_{\mp n} s_{\mp n} \b_2(\mp n)\right] 
\eqend
where $J^+(\al)$ and $J^-(\al)$ are the creation-and annihilation parts of 
$J(\al)$, i.e., $J^-(\al)\Omega = \, J^+(\al)^*\Omega =0$. 
Moreover, the forms $\Phi$ with $\alpha_n$ such that
\eq
\label{loopcond}
\sum_{n\in\Z}|n||\alpha_n|^2<\infty
\eqend
generate a $*$-algebra of forms on $\cD$, with
\eq
\label{PHIs}
(\xxa R^w \ee{\ii J(\al)}\xxe)^* = \, \xxa R^{-w} \ee{-\ii J(\al)^*}\xxe\, , 
\quad  J(\al)^* = \sum_{n\in\Z}\overline{\alpha_{-n} }\b(-n)
\eqend
and the multiplication rule obtained with the Hausdorff formula
\eq
\label{Hausdorff}
\ee{A}\ee{B} = \ee{ c/2 } \ee{A+B} = \ee{ c } \ee{B}\ee{A}\quad 
\mbox{ if $[A,B]=c\id$, $c\in\C$} 
\eqend
and Eqs.\ \Ref{12}--\Ref{QRQ}, i.e., 
\eqa
\label{mult}
\xxa  R^w \ee{\ii J(\al)} \xxe \xxa  R^{w'} \ee{\ii  J(\be)} \xxe = 
\ee{- [ J^-(\al),  J^+(\be)] + \ii ( \alpha_0 w' -\beta_0 w )/2 }
\xxa R^{w+w'}\ee{ \ii J(\al+\be) } \xxe 
\eqaend
where
\eq
\label{JmJp}
[ J^-(\al),  J^+(\be)] = \sum_{n=1}^\infty n\left( 
|c_n|^2\alpha_n\beta_{-n} + 
|s_n|^2 \alpha_{-n}\beta_{n}\right) \, . 
\eqend
Moreover, 
\eq
\label{w0}
\pr{\Omega}{ \xxa R^w\ee{\ii  J(\al) }\xxe  \Omega} = \delta_{w,0} \: . 
\eqend
\end{Lem}

\begin{Rem}\label{2.3}
As discussed in Appendix~B, 
for $\alpha_n$ obeying the conditions in Eq.\ \Ref{loopcond1}, 
the forms $R^w\ee{\ii  J(\al)}$ define unitary operators and 
provide a unitary representation of the loop group of $\U(1)$ on $\cF$. Note 
that the Hausdorff formula 
implies
\eq
\label{hm}
\xxa\ee{\ii  J(\al)}\xxe \, = 
\ee{-[ J^-(\al), J^+(\al)]/2} \ee{\ii  J(\al)}\: , 
\eqend
i.e., the forms $\Phi$ in this case are proportional to
unitary operators. We will not make use 
of these facts in this paper and will regard such  
$\Phi$'s only as sesquilinar forms. 
\end{Rem}

\begin{Rem}\label{2.4}
Eq.\ \Ref{JmJp} explains the condition in Eq.\ 
\Ref{cnsn3} above: recalling that $|c_n|^2=1+|s_n|^2$ and 
Cauchy's inequality, we see that 
this is the natural condition ensuring that $[ J^-(\al),  J^+(\be)]$
is well-defined for all $\al$ and $\be$ obeying the condition in Eq.\ 
\Ref{loopcond}.
\end{Rem}

\noindent {\it Proof of Lemma~\ref{L0}:}

The l.h.s.\ of Eq.\ \Ref{mult} equals 
\nonueqa
\ee{\ii\alpha_0 Q/2 }R^w\ee{\ii\alpha_0 Q/2}\ee{\ii  J^+(\al)}\ee{\ii  J^-(\al)}
\ee{\ii\beta_0 Q/2 } R^{w'} \ee{\ii\beta_0 Q/2}\ee{\ii  J^+(\be)}\ee{\ii  J^-(\be)} 
= \nonu = 
\ee{-\ccr{ J^-(\al)}{ J^+(\be)} + \ii ( \alpha_0 w' -\beta_0 w )/2 }
\ee{\ii(\alpha_0+\beta_0) Q/2 }R^{w+w'}
\ee{\ii(\alpha_0 +\beta_0)Q/2}\nonu \times 
\ee{\ii ( J^+(\al)+ J^+(\be))}\ee{\ii ( J^-(\al) + J^-(\be))} 
\nonueqaend
which obviously is equal to the r.h.s.\ of Eq.\ \Ref{mult}. \QED

\subsection{Construction of anyons}

\noindent {\bf Definition:} {\it 
The {\it anyons} associated with a real, non-zero 
parameter $\nu$ are, 
\eq
\label{phi}
\phi^\mu_\eps(x)  \, : = \,   
\ee{-\ii\mu\nu^2 Qx/2}\, R^\mu\, \ee{- \ii\mu\nu^2 Qx/2} 
\xxa \ee{ - \ii\mu\nu K_\eps(x) } \xxe\: , \quad \mu\in\Z 
\eqend
with 
\eq
\label{K}
K_\eps(x) \, := \, \sum_{n\in\Zp} \f{1}{\ii n}\, \ee{\ii n x -|n|\eps}
\b(n) \eqend where $-\pi\leq x \leq \pi$ is a coordinate on the circle
and $\eps>0$ a regularization parameter. In particular, we define
\eq
\label{m1}
\phi_\eps(x)\, := \, \phi_\eps^{1}(x) \,  .  
\eqend 
}

\bigskip
 
The main properties of these anyons are summarized in 
the following

\begin{Prop}\label{P1}
The $\phi^\mu_\eps(x)$ 
generate a $*$-algebra of sesquilinar forms on $\cD$ 
and obey 
\eq
\label{adj}
\phi^\mu_\eps(x)^* = \phi^{-\mu}_\eps(x)
\eqend
and the exchange relations
\eq
\label{exc}
\phi^\mu_\eps(x)\phi^{\mu'}_{\eps'}(y) =
\ee{-i\pi\mu\mu'\nu^2 \sgn_{\eps+\eps'}(x-y)}
\phi^{\mu'}_{\eps'}(y)\phi^\mu_\eps(x)  
\eqend
where
\eq
\sgn_\eps(x) \, := \, \frac{1}{\pi} \left(x+ 
 \sum_{n\in \Zp } 
\f{1}{\ii n} \, \ee{\ii n x -|n|\eps} \right)    
\eqend 
a regularized sign function on the circle, i.e., 
it is $C^\infty$ and  converges to $\sgn(x)$ 
in the limit $\eps\downarrow 0$. Moreover, 
\eq
\label{corrfun}
\pr{\Omega}{ \phi_{\eps_1}^{\mu_1}(x_1) \cdots    \phi_{\eps_M}^{\mu_1}(x_M)\Omega} = 
\delta_{\mu_1+\cdots+\mu_N,0}
 \prod_{1\leq j<k\leq M} b_{\eps_j+\eps_k}(x_j-x_k)^{\mu_j\mu_k\nu^2 } 
\eqend
with 
\eq
\label{beps}
b_\eps(r) \, := \, \ee{-\ii r/2 - C_\eps(r) } \, , \quad  
C_\eps(r)\, := \, \sum_{n=1}^\infty 
\f{1}{n}\left(  |c_n|^2 \ee{\ii nr -n\eps}
+ |s_n|^2 \ee{-\ii n r-n\eps} \right) \: . 
\eqend
In particular, for $c_n$ and $s_n$ as in Eq.\ \Ref{expl}, 
\eq
\label{b1}
b_\eps(x)\, :=\, 2\ee{3\pi\ii/2 - \eps/2}\sin\half(x+i\eps) 
\prod_{n=1}^\infty (1-2q^{2n}\ee{-\eps}\cos(x)+q^{4n}\ee{-2\eps} )\: .  
\eqend
\end{Prop}

\begin{Rem}\label{1}
The relations in Eq.\ \Ref{exc} show that the $\phi_\eps(x)$ are
regularized anyons with statistics parameter \eq
\label{m2}
\lambda=\nu^2 
\eqend
as described in the Introduction. 
\end{Rem}

\begin{Rem}\label{2}
We stress that we introduce 
the parameter $\eps>0$ as a convenient technical tool:
the $\phi^\mu_\eps(x)$ are {\it regularized quantum fields},  
i.e., in the limit $\eps\downarrow 0$ they become 
operator valued distributions whereas for $\eps>0$ they
are well-defined operators and, in particular, 
can be multiplied with each other. Indeed, it follows from
Remark \ref{2.3} above that $\phi^\mu_\eps(x)$ is proportional
to the unitary operator 
$$
\ee{-\ii\mu\nu^2 Qx/2}\, R^\mu\, \ee{- \ii\mu\nu^2 Qx/2} 
 \ee{ - \ii\mu\nu K_\eps(x) } 
$$
with the proportionality constant $b_{2\eps}^{-(\mu\nu)^2/2}$ which diverges
in the limit $\eps\downarrow 0$. 
Eventually we are interested in this singular 
limit $\eps\downarrow 0$, but we will be able to take 
this limit at a later point without difficulty.

\end{Rem}

\noindent {\em Proof:} The proof is by straightforward computations
using the Lemma~\ref{L0} above. In particular, Eq.\ \Ref{adj} is 
trivial consequences of Eq.\ \Ref{PHIs}, and 
to prove Eq.\ \Ref{exc} it is sufficient 
to note the following identity,  
$$
\ccr{K_\eps(x)}{K_{\eps'}(y)} = 
\sum_{n\in \Zp }\f{1}{n}\, \ee{\ii n (x-y) -|n|(\eps+\eps') }\id  \: , 
$$
Eq.\ \Ref{QRQ} and the Hausdorff formula Eq.\ \Ref{Hausdorff}. 
To see that $\sgn_\eps(x)$ is a regularized sign function one only needs 
to check that it is an odd function in $x$, and its $x$-derivative 
divided by 2 
equals
\eq
\label{delta}
\delta_\eps(x) \, := \, \frac{1}{2\pi}  \sum_{n\in \Z} \ee{ \ii n x -|n|\eps}     
\eqend
which obviously is a regularized delta function.

To compute the normal ordering of products of anyons
we determine the creation- and annihilation parts $K_\eps^\pm(x)$ 
of the form $K_\eps(x)$, 
\eq
\label{Kpm}
K_\eps^\pm(x) = \mp\sum_{n=1}^\infty
 \f{1}{\ii n}\left( c_{\pm n}\, \ee{\mp \ii nx-n\eps} \b_1(\mp n) - 
  s_{\mp n} \, \ee{\pm \ii nx-n\eps} \b_2(\mp n) \right) 
\eqend
and compute 
\eq
\label{KpKm}
\ccr{K^-_\eps(x)}{K^+_{\eps'}(y)} = 
C_{\eps+\eps'}(x-y)\id
\eqend
with $C_{\eps}(r)$ given in Eq.\ \Ref{beps}. 
The normal ordering of an arbitrary product of anyons
is defined as follows, 
\nonueqa
\xxa \phi_{\eps_1}^{\mu_1}(x_1) \cdots    \phi_{\eps_M}^{\mu_M}(x_M)  \xxe 
 \, =\,  
\ee{-\ii\nu^2 Q(\mu_1 x_1 + \cdots + \mu_M x_M ) /2}\, R^{\mu_1+\ldots + \mu_M} \, 
\ee{- \ii \nu^2 Q(\mu_1 x_1 + \cdots + \mu_M x_M ) /2} \nonu \times
\ee{ -\ii\nu [ \mu_1 K^+_{\eps_1}(x_1)+ \cdots + \mu_M K^+_{\eps_M}(x_M)]  }
\ee{ -\ii\nu [ \mu_1\nu K^-_{\eps_1}(x_1)+ \cdots + 
\mu_M K^-_{\eps_M}(x_M) ] } \, , 
\nonueqaend
and by repeated application of Eq.\ \Ref{mult} we obtain
\eqa
\label{nocf}
 \phi_{\eps_1}^{\mu_1}(x_1) \cdots    \phi_{\eps_M}^{\mu_M}(x_M) 
=   \prod_{1\leq j<k\leq M} b_{\eps_j+\eps_k}(x_j-x_k)^{\mu_j\mu_k\nu^2}
\xxa \phi_{\eps_1}^{\mu_1}(x_1) \cdots    \phi_{\eps_M}^{\mu_M}(x_M)  \xxe 
\eqaend
with $b_\eps(r)$ in Eq.\ \Ref{beps}.
This and Eq.\ \Ref{w0} prove Eq.\ \Ref{corrfun}. 
Eq.\ \Ref{b1} is proven in Appendix~A. 

\QED

\begin{Rem}\label{R0}
In computations it is sometimes convenient to write
\eq
\phi_\eps^\mu(x) =\,  \xxa \ee{-\ii \nu\mu \cX_\eps(x) } \xxe
\eqend
with
\eq
\cX_\eps(x) = q + px + \sum_{n\in\Zp} \f{1}{\ii n}\, \ee{\ii n x -|n|\eps} \b(n)
\eqend
with $q=\nu Q$ and $p = `` (\ii/\nu)\log(R)$'' formally obeying
\eq
\ccr{q}{p}=\ii\id\, , \quad \ccr{p}{\b(n)}=\ccr{q}{\b(n)}=0 \quad 
\forall n\neq 0\: , 
\eqend
and therefore $\ccr{\cX_\eps(x)}{\cX_{\eps'}(y)} = -
\ii\pi\sgn_{\eps+\eps}(x-y)$.  This makes many computations simpler,
but it is important to remember that $p$ itself is not well-defined
but only its exponentials $\ee{-\ii\mu\nu p}\equiv R^\mu$ and only for
integers $\mu$. This makes clear that our anyons are essentially
vertex operators used, e.g., in string theory (see e.g.\
\cite{strings}).
\end{Rem}

Proposition~\ref{P1} above summarizes the main properties of anyons.
In particular, it gives explicit formulas for all so-called {\em anyon 
correlation functions} which are defined by the l.h.s.\ of Eq.\ 
\Ref{corrfun} and 
which, according to physics folklore, capture the physical properties
of the anyon system. For the eCS model the following anyon correlation
function will play a central role, 
\eq
\label{F1}
F^{\eps',\eps}_N(\vx ; \vy)
\,  : = \,   
\pr{\Omega}{ \phi_{\eps}(x_N)^*
\cdots \phi_{\eps}(x_1)^* \phi_{\eps'}(y_1)\cdots 
\phi_{\eps'}(y_N)\Omega} \: ,   
\eqend
and with Proposition~\ref{P1} we obtain the following explicit formula, 
\eq
\label{F}
F^{\eps,\eps'}_N(\vx ; \vy)
=
\f{ \prod_{1\leq j<k\leq N} b_{2\eps}(x_{k}-x_j)^{\lambda} 
\prod_{1\leq j<k\leq N} 
b_{2\eps'}(y_{j}-y_{k})^{\lambda}}{\prod_{j,k=1}^N
b_{\eps+\eps'}(x_j-y_k)^{\lambda}}    
\eqend
which will be important for us later on.

\section{Second quantization of the eCS model}
In this Section we construct the second quantization of the 
eCS model, i.e., find a self-adjoint form $\cH$ on $\cF$ 
such that the commutator of $\cH$ with a product of anyons, 
\eq
\label{PhiN}
\Phi^{N}_\eps (\vx)\; := \, 
\phi_{\eps}(x_1)\cdots \phi_{\eps}(x_N)\, , 
\eqend
$\phi_\eps(x)$ defined in Eqs.\ \Ref{K}--\Ref{m1}, 
is essentially equal to the eCS Hamiltonian applied to 
this product. To be precise

\begin{Prop}\label{P2} 
Let  $\b(n)$ be as in Eqs.\ \Ref{bn}--\Ref{cnsn2} and 
\eqa
\label{cH}
\cH \, := \, \nu W^3 + (1-\nu^2)\cC 
+ 2\nu (\nu-1) W^2 Q + \f{1}{3}\nu(\nu-1)^2(2+\nu) Q^3 - \nu^4 c_\eps Q 
\eqaend
with 
\eqa
\label{Wdef}
W^2 &:=& \f{1}{2}\sum_{n\in\Z} \xxa\b(-n)\b(n)\xxe \nonu
W^3 &:=&  \f{1}{3} \sum_{m,n\in\Z} \xxa \b(-m)\b(-n)\b(m+n) \xxe  \nonu 
\cC &:=& \sum_{n=1}^\infty n\left[\b_1(-n)\b_1(n) + \b_2(-n)\b_2(n) \right]   
\eqaend
and the constant
\eq
\label{cceps}
c_\eps =   \f{1}{12}  -  2\sum_{n=1}^\infty n|s_n|^2 \ee{-2|n|\eps} 
-  
\sum_{m,n=1}^\infty 2 |s_m|^2|s_n|^2\ee{-(m+n)\eps}\left(\ee{-|m-n|\eps} - 
\ee{-|m+n|\eps} \right)  \: . 
\eqend
Then 
\eq
\label{sq1}
\ccr{\cH}{\Phi^{N}_\eps (\vx)}\Omega = H_N^{2\eps}\Phi^{N}_\eps (\vx)\Omega + 
\eps \nu \Psi^N_\eps(\vx)\Omega
\eqend
where
\eq
\label{reCS}
H^\eps_{N} \, := \,  - \sum_{j=1}^N\frac{\partial^2}{\partial x_j^2} 
\; + \;   2 \nu^2 (\nu^2-1)\!\! \sum_{1\leq j<k\leq N}  V_\eps(x_j-x_k)
\eqend
with
\eq
\label{Veps}
V_\eps(r) \, := \, 
- \sum_{n=1}^\infty n\left(  |c_n|^2 \ee{\ii nr -n\eps} 
+ |s_n|^2 \ee{-\ii n r-n\eps} \right) 
\eqend
is a regularized many particle Hamiltonian, and 
\eq
\label{Psi}
\Psi_\eps^N(\vx) \, := \, \sum_{j=1}^N \phi_\eps(x_1)\ldots \phi_\eps(x_{j-1})
\xxa  \phi_\eps(x_j)\rr_\eps (x_j)\xxe \phi_\eps(x_{j+1}) \cdots \phi_\eps(x_N) 
\eqend
with 
\eqa
\label{rr} 
\rr_\eps(x) \, : = \, \sum_{m,n=1}^\infty 2 \nu \left[ \nu |s_m|^2 
\left( \b(n)\ee{\ii nx} -  \b(-n)\ee{-\ii nx} \right) 
-  \xxa \b(-m)\b(n) \xxe \right]
\nonu \times  
\f{1}{\eps}\left(\ee{-|m-n|\eps} - \ee{-|m+n|\eps} \right)\ee{-|m|\eps} \: . 
\eqaend
In particular, if  $c_n$ and $s_n$ are as in Eq.\ \Ref{expl} then
\eq
\label{V2}
V_\eps(r) = -\f{\partial^2}{\partial r^2} \log\left(
\sin\half(r+\ii\eps)\prod_{m=1}^\infty
\left[1-2q^{2m}\ee{-\eps}\cos(r)+q^{4m} \ee{-2\eps} \right]\right) \,
, \eqend i.e., $H^\eps_{N}$ is a regularized version of the eCS
differential operator in Eq.\ \Ref{eCS} with
$\lambda=\nu^2$. Moreover, if $c_n$ and $s_n$ are as in Eq.\
\Ref{expl} then also the following identity holds true, 
\eq
\label{sq2}
\pr{\Omega}{\ccr{\cH}{\Phi^{N}_{\eps} (\vx)^* \Phi^{N}_{\eps'} (\vy)}
\Omega} = 0 \: .   
\eqend
\end{Prop}

\begin{Rem}\label{1.1}
Note that the sesquilinear form $\rr_\eps(x)$ in Eq.\ \Ref{rr} is
non-singular in the limit $\eps\downarrow 0$, hence the
second term on the r.h.s.\ of Eq.\ \Ref{sq1} vanishes in this limit. 
It is therefore natural to write Eq.\ \Ref{sq1} in the following 
suggestive form
\eq
\label{sq11}
\ccr{\cH}{\Phi^{N}_\eps (\vx)}\Omega \simeq  H_N^{2\eps}\Phi^{N}_\eps (\vx)\Omega
\eqend
where the symbol `$\simeq$' means `equal in the limit $\eps\downarrow 0$'.
\end{Rem}
 
\begin{Rem}\label{1.2}
As we will see, this proposition implies a remarkable identity
(Proposition~\ref{P3} in the next section) which is the starting point
of our solution algorithm for the eCS model. We note that Eq.\
\Ref{sq1} comes for free in the Sutherland case $q=0$ where
$\cH\Omega=0$, but this is no longer holds true in general. It is also
interesting to note that a form $\cH$ obeying the relations in Eq.\
\Ref{sq1} exists for a much larger class of models: in the first part
of this proposition we do not need to assume Eq.\ \Ref{expl}.
However, to prove Eq.\ \Ref{sq2} we need Eq.\ \Ref{expl}: it is this
relation which restricts this proposition to interaction potentials
which are equal to Weierstrass' $\wp$-function.
\end{Rem}

\begin{Rem}\label{1.3}
We note that only the first two terms on the r.h.s.\ of Eq.\ \Ref{cH}
are really important, and dropping the other terms can be compensated
by adding trivial terms to the eCS differential operator: simpler
variants of our proof of Lemma~\ref{L1} yield \eqa
\ccr{Q}{\Phi_\eps^N(\vx)} &=& N\Phi_\eps^N(\vx) \nonu \ccr{W^2 +
\half(\nu-1)(\nu-3)Q^2}{\Phi_\eps^N(\vx)} &=& \sum_{j=1}^N
\ii\f{\partial}{\partial x_j} \Phi_\eps^N(\vx) \, .  \eqaend This
shows that we have a relation \eq \ccr{\nu
W^3+(1-\nu^2)\cC}{\Phi^N_\eps(\vx)}\Omega \simeq \tilde H^{2\eps}_N
\Phi^N_\eps(\vx)\Omega \eqend where $\tilde H^{2\eps}_N$ equals
$H^{2\eps}_N$ up to a constant and a multiple of the total momentum
operator $P_N\, := \, -\sum_{j=1}^N \ii \partial/ \partial x_j $.
Note also that the change $\cH\to \nu W^3+(1-\nu^2)\cC$ does not
affect the identity in Eq.\ \Ref{sq2} since $Q\Omega= W^2\Omega=0$.

\end{Rem}

\noindent {\em Proof of Proposition~\ref{P1}:} 
We prove this proposition in several steps. We start by
stating two Lemmas which provide the essential properties
of the building blocks of $\cH$. The proofs of
these Lemmas are by straightforward but tedious computations 
which we defer to Appendix~C.

\begin{Lem}\label{L1}
The sesquilinear form
\eq
\label{Wtdef}
\tW^3 \, := \, \lim_{\eps \downarrow 0}
\f{1}{6\pi }\int_{-\pi}^{\pi} dx\, \xxa \bb_\eps(x)^3 \xxe  -\nu^3 c_\eps Q
\eqend
with 
\eq
\label{bb}
\bb_\eps(x) = \nu Q +\sum_{n\in\Zp} \b(n) \ee{\ii nx-|n|\eps} 
\eqend
and $c_\eps$ defined in Eq. \Ref{cceps}, obeys the following relations
\eqa
\label{Wt3}
\ccr{\tW^3}{\phi_\eps(x)} \, &=& \, 
- \f{1}{\nu}\f{\partial^2}{\partial x^2} \phi_\eps(x)      
+ \ii(\nu^2 - 1) \xxa 
\bb'_\eps(x) \phi_\eps(x)\xxe + \eps \xxa \rr_\eps(x)\phi_\eps(x)\xxe 
\eqaend
with $\rr_\eps(x)$ defined in Eq.\ \Ref{rr} 
and $\bb'_\eps(x) = \partial\bb_\eps(x)/\partial x$. 
\end{Lem}

\noindent {\it (Proof in Appendix~C.1)}
 
\begin{Rem}\label{1.4}
It is useful to note
that $\tW^3$ accounts for all but one term of $\cH$: 
the forms $W^s$ ($s=2,3$) defined 
in Eq.\ \Ref{Wdef} can also be computed as follows, 
$W^s =  \lim_{\eps \downarrow 0}
\int_{-\pi}^{\pi} dx\, \xxa \rho_\eps(x)^s \xxe/(2\pi s)$,  
with $\rho_\eps(x) = \sum_{n\in\Z} \b(n)\ee{\ii nx-|n|\eps}$. 
Since $\bb_\eps(x) = \rho_\eps(x) + (\nu-1)Q$, this implies
\eqa
\tW^3 &=& W^3 + 2(\nu-1) W^2 Q + \f{1}{3}(\nu-1)^2 (\nu+2) Q^3 
- \nu^3 c_\eps Q  \: , 
\eqaend
which shows that $\cH$ in Eq.\ \Ref{cH} is identical with
\eq
\cH = \nu\tW^3 + (1-\nu^2)\cC \: . 
\eqend

\end{Rem}

\begin{Lem}\label{L2}
For all sesquilinear forms 
$\Phi$ as in Eq.\ \Ref{PHI}, 
the sesquilinear form $\cC$ defined in Eq.\ \Ref{Wdef} obeys
\eq
\label{PHI1}
\cC \Phi + \Phi \cC = 2\xxa  \Phi\cC \xxe - \ii \xxa  J(\al'') 
\Phi\xxe\: , \quad
\mbox{ with } 
J(\al'') \, := \,  -\sum_{n\in\Z} n^2 \alpha_n\b(-n) \: .  
\eqend
\end{Lem}

\noindent {\it (Proof in Appendix~C.2)}

\bigskip 

In particular, this Lemma implies 
\eqa
\label{PHI2}
\cC\xxa \Phi^N_\eps(\vx)\xxe + \xxa \Phi^N_\eps(\vx)\xxe \cC  \, =
2 \xxa \Phi^N_\eps(\vx) \cC \xxe +  
\sum_{j=1}^N \ii \nu \xxa \bb_\eps'(x_j)\Phi^N_\eps(\vx)\xxe \: . 
\eqaend
To see that, recall that
$
\xxa \Phi^N_\eps(\vx)\xxe\,  =\, \xxa R^N  \ee{\ii J(\al) } \xxe 
$
with
$$
J(\al) =  
-\nu \sum_{j=1}^N \left( \nu x_j Q + \sum_{n\in\Zp}\f{1}{\ii n}\b(n)\ee{\ii nx_j-|n|\eps} 
\right) \: , 
$$
and therefore in this case
$$
J(\al'')=  -\nu \sum_{j=1}^N  
\sum_{n\in\Zp} \ii n \b(n)\ee{\ii nx_j-|n|\eps} \equiv  
-\nu \sum_{j=1}^N  \bb_\eps'(x_j) \: .  
$$

We now show how Lemmas \ref{L1} and Eq.\ \Ref{PHI2} imply Eq.\
\Ref{sq1}.  From the definition in Eqs.\ \Ref{PhiN} we get by a simple
computation, using repeatedly Eq.\ \Ref{Wt3},
\nonueqa
\ccr{\tW^3 }{\Phi_\eps^N(\vx)} = 
\sum_{j=1}^N \phi_\eps(x_1)\cdots \ccr{\tW^3}{\phi_\eps(x_j)}\cdots  \phi_\eps(x_N) 
= - \f{1}{\nu}\sum_{j=1}^N\f{\partial^2}{\partial x_j^2} \Phi_\eps(\vx) \nonu +
\ii (\nu^2 - 1) \sum_{j=1}^N\phi_\eps(x_1)\cdots \xxa 
\bb'_\eps(x_j) \phi_\eps(x_j)\xxe   \cdots  \phi_\eps(x_N) 
 + \eps \Psi^N_\eps(\vx)
\nonueqaend
with $\Psi^N_\eps(\vx)$ given in Eq.\ \Ref{Psi}.
Moreover, Eq.\ \Ref{PHI2} together with 
\eq
\Phi_\eps(\vx) = \Delta^{2\eps}_N(\vx) \xxa\Phi_\eps(\vx)  \xxe \: , 
\quad \Delta^{2\eps}_N(\vx)\,   :=  \, 
\prod_{1\leq j<k\leq N} b_{2\eps}(x_j-x_k)^{\nu^2}  
\eqend
(which is a special case of Eq.\ \Ref{nocf}) and the definition 
of normal ordering imply 
\nonueqa
\ccr{\cC}{ \Phi_\eps(\vx) } = 
-2\Phi^N_\eps(\vx)\cC + \Delta^{2\eps}_N(\vx) \left( 
2 \xxa \Phi^N_\eps(\vx)\cC \xxe +  
\sum_{j=1}^N \ii \nu \xxa \bb_\eps'(x_j)\Phi^N_\eps(\vx)\xxe 
\right) \nonu
= -2\Phi^N_\eps(\vx)\cC + 2 \Delta^{2\eps}_N(\vx) 
\xxa \Phi^N_\eps(\vx)\cC \xxe
+ 
\sum_{j=1}^N \ii \nu \left[  (\bb_\eps')^-(x_j)\Phi^N_\eps(\vx) + 
\Phi^N_\eps(\vx) (\bb_\eps')^-(x_j)\right]  
\nonueqaend
with $(\bb')^\pm _\eps(y) = \partial \bb_\eps^\pm(y)/\partial y = 
\partial^2 K_\eps^\pm(y)/\partial y^2 $ the creation- and annihilation
parts of $\bb_\eps'(y)$. We thus get
\eqa
\label{abc}
\ccr{\cH}{\Phi^N_\eps(\vx)}\Omega = \nu \ccr{\tW^3 }{\Phi_\eps^N(\vx)}\Omega 
+ (1-\nu^2)\ccr{\cC}{ \Phi_\eps(\vx) }\Omega \nonu = 
\left[ - \sum_{j=1}^N\f{\partial^2}{\partial x^2_j} \Phi_\eps(\vx)  +
\ii \nu (\nu^2 - 1) (\cdot) + 
\eps \nu \Psi^N_\eps(\vx) \right]\Omega
\eqaend
with 
$$
(\cdot) \, := \, 
\sum_{j=1}^N\left[  
\phi_\eps(x_1)\cdots \xxa \bb'_\eps(x_j) \phi_\eps(x_j)\xxe   \cdots  \phi_\eps(x_N) - 
(\bb_\eps')^-(x_j)\Phi^N_\eps(\vx) - \Phi^N_\eps(\vx) (\bb_\eps')^-(x_j) \right]  
$$
where we used $\cC\Omega=\, \xxa\Phi^N_\eps(\vx)\cC\xxe\Omega  = 0$.
We now recall Eq.\ \Ref{KpKm} which implies 
$$
\ccr{(\bb')^\mp_\eps(y)}{K^\pm_{\eps'}(x)} = 
\f{\partial^2}{\partial y^2}\ccr{K^\mp_\eps(y)}{K^\pm_{\eps'}(x)} = 
\pm \f{\partial^2}{\partial y^2} C_{\eps+\eps'}( \mp (x-y) )\id \: , 
$$
and with Eqs.\ \Ref{beps} and \Ref{Veps}, 
$$
\ccr{(\bb')^\mp_\eps(y)}{K^\pm_{\eps'}(x)} =  
\pm V_{2\eps}( \mp(x-y) ) \id  \: . 
$$
With that we obtain 
$$
\ccr{(\bb')^\mp_\eps(y)}{\phi_\eps(x)} = 
-\ii\nu \ccr{(\bb')^\mp_\eps(y)}{K^\pm_\eps(x)} \phi_\eps(x)
= \mp \ii\nu  V_{2\eps}( \mp(x-y) ) \phi_\eps(x) \: , 
$$
and with $\bb'_\eps(x_j)=  (\bb')^+_\eps(x_j) + (\bb')^-_\eps(x_j)$ 
and repeated application of these latter identities we get 
\nonueqa
\phi_\eps(x_1)\cdots \xxa 
\bb'_\eps(x_j) \phi_\eps(x_j)\xxe   \cdots  \phi_\eps(x_N)
= (\bb')^+_\eps(x_j)\Phi_\eps^N(\vx) + 
\Phi_\eps^N(\vx) (\bb')^-_\eps(x_j)\nonu
- \ii\nu \left[ \sum_{k=1}^{j-1} V_{2\eps}(x_k-x_j ) + 
\sum_{k=j+1}^{N}V_{2\eps}(x_j-x_k)  \right] \Phi_\eps^N (\vx) \, ,  
\nonueqaend
and thus 
\nonueqa
(\cdot) = 
-\ii \nu \sum_{j=1}^N \left[ \sum_{k=1}^{j-1} V_{2\eps}(x_k-x_j ) + 
\sum_{k=j+1}^{N}V_{2\eps}(x_j-x_k)  \right] \Phi_\eps^N (\vx) \nonu 
\equiv 
- 2 \ii \nu \sum_{1\leq j<k\leq N} V_{2\eps}(x_j-x_k)\Phi_\eps^N(\vx) \: . 
\nonueqaend
If we insert this in Eq.\ \Ref{abc} we obtain Eqs.\ \Ref{sq1}--\Ref{reCS}. 

Finally, we note that validity of
Eq.\ \Ref{sq2} under the condition Eq.\ \Ref{expl}
is obviously a special case of the following more general
result:

\begin{Lem}\label{L3}
For all sesquilinear forms $\Phi$ 
as in Eq.\ \Ref{PHI} and for the form $\cH$ defined in Eqs.\ 
\Ref{cH}--\Ref{Wdef}, the 
relation
\eq
\label{sq21}
\pr{\Omega}{\ccr{\cH}{\Phi}\Omega}=0
\eqend
holds true provided that
\eq
\label{crucial}
|c_m|^2|c_n|^2|s_{m+n}|^2 = |s_m|^2|s_n|^2|c_{m+n}|^2 \quad \forall m,n\in\N \: . 
\eqend
\end{Lem}

\noindent {\it (Proof in Appendix~C.3)}

\begin{Rem}\label{1.7}
Eq.\ \Ref{sq21} is non-trivial only for $w = 0$. 
\end{Rem}

This completes the proof of Proposition~\ref{P2}. \QED
 
\begin{Rem}\label{1.6}
It is obvious that $c_n$ and $s_n$ as in Eq.\ \Ref{expl} fulfill the
condition in Eq.\ \Ref{crucial}. It is interesting to note that Eq.\
\Ref{expl} actually accounts for all possibilities: simple
computations shows that the conditions in Eqs.\ \Ref{cnsn1} and
\Ref{crucial} are equivalent to:
$$
Q_n \, := \, 1-\f{1}{|c_n|^2} \quad \mbox{ obey } 
\quad Q_{m+n} = Q_m Q_n \quad \forall m,n\in\N \: , 
$$
and all solutions of the latter are of the form $Q_n=q^{2n}$ for some complex
$q$. If we also require Eq.\ \Ref{cnsn2} then $q^2$ must be real, and
Eq.\ \Ref{cnsn3} further restricts to $|q^2|<1$.  

\end{Rem}

\section{A remarkable identity}
From the results in the previous Section we now obtain the following 

\begin{Prop}\label{P3} The anyon correlation function
in Eq.\ \Ref{F} obeys the following identity,
\eq
\label{rem}
\biggl[
\overline{H_N^{2\eps}(\vx)} - H_N^{2\eps'}(\vy)
\biggr] 
F^{\eps,\eps'}_N(\vx ; \vy) = R^{\eps,\eps'}(\vx;\vy)
\eqend
where the regularized eCS differential operators are defined in Eqs.\ 
\Ref{reCS}--\Ref{Veps} and act on
different arguments as indicated, and 
\eq 
R^{\eps,\eps'}(\vx;\vy) = -\eps'
\pr{\Omega}{\Phi^N_{\eps}(\vx)^*\Psi^N_{\eps'}(\vy)\Omega} + 
\eps\pr{\Omega}{\Psi^N_{\eps}(\vx)^*\Phi^N_{\eps'}(\vy)\Omega}\, , 
\eqend 
$\Psi^N_\eps(\vx)$ defined in Eqs.\ 
\Ref{Psi}--\Ref{rr}, are correction terms vanishing pointwise in the limit 
$\eps,\eps'\downarrow 0$. 
\end{Prop}

\begin{Rem} Note that the complex conjugation in Eq.\ \Ref{rem}
is important as it affects the regularization: since
$\overline{V_\eps(r)} =V_\eps(-r)$ it amounts to replacing the
regularized singularity $\sim 1/(r+\ii\eps)^2$ by $\sim
1/(r-\ii\eps)^2$.
\end{Rem} 
 
\begin{Rem} One can express this result as follows,  
\eq
\label{rem1}
\overline{H_N^{2\eps} (\vx)}F^{\eps,\eps'}_N(\vx ; \vy) \simeq  
H_N^{2\eps'}(\vy)F^{\eps,\eps'}_N(\vx ; \vy) \, .   
\eqend 
\end{Rem}

\begin{Rem} In the limits $\eps,\eps'\downarrow 0$ the function $F^{\eps,\eps'}_N(\vx ; \vy)$
becomes equal, up to a constant, to the function
\eq
F_N(\vx;\vy) =
\f{ \prod_{1\leq j<k\leq N} \tet(x_{k}-x_j)^{\lambda} 
\prod_{1\leq j<k\leq N} 
\tet(y_{j}-y_{k})^{\lambda}}{\prod_{j,k=1}^N
\tet(x_j-y_k)^{\lambda}}    
\eqend
with $\tet(r)$ in Eq.\ \Ref{tet}, and we obtain the following identity,
\eq
\sum_{j=1}^N\biggl(\f{\partial^2}{\partial x_j^2} -  
\f{\partial^2}{\partial y_j^2}  \biggr)F_N(\vx;\vy) 
=  2\lambda(\lambda-1)\sum_{1\leq j<k\leq N}\biggl( 
V(x_k-x_j) - V(y_j-y_k) \biggr)F_N(\vx;\vy)  
\eqend
with $V(r)$ in Eq.\ \Ref{V}. Recalling that $\tet(r) = const.\ 
\vartheta_1(r/2)$
and $V(r) = \wp(r)+ const.$ we see that this is a remarkable identity for  
elliptic functions. 
\end{Rem}

\noindent {\em Proof of Proposition~\ref{P3}:} 
We start with the following (trivial) identity
\nonueqa
[\cH, \Phi^N_{\eps}(\vx)^* \Phi^N_{\eps'}(\vy)] = 
\Phi^N_{\eps}(\vx)^*[\cH,\Phi^N_{\eps'}(\vy)]- [\cH, \Phi^N_{\eps}(\vx)]^* \Phi^N_{\eps'}(\vy)
\nonueqaend
(where we used $\cH=\cH^*$)
and compute its vacuum expectation value using Eq.\ \Ref{sq2},
\nonueqa
\pr{ \Omega}{\Phi^N_{\eps}(\vx)^*[\cH,\Phi^N_{\eps'}(\vy)] \Omega} 
- \pr{\Omega}{ [\cH, \Phi^N_{\eps}(\vx)]^*\Phi^N_{\eps'}(\vy) \Omega} = 0 \, . 
\nonueqaend
Inserting now Eqs.\ \Ref{sq2} and recalling Eqs.\ \Ref{F1}--\Ref{F}
we obtain the result. \QED

\section{Conclusion} As stated in the beginning of Section~1, 
the results of this paper provide the starting point for an algorithm
to construct eigenfunctions and eigenvalues of the eCS model, and this
algorithm will be elaborated in \cite{EL2}.  It is interesting to note
that in the limiting case $q=0$ this algorithm is different from the
one of Sutherland, even though it is equivalent to it in the sense
that it yields the same solutions and is equally simple (a detailed
comparison of these algorithms is given in \cite{EL3}).

For the convenience of the reader we outline here how Proposition
\ref{P3} is used to obtain this algorithm. The idea is to take the
Fourier transform of the identity in Eq.\ \Ref{rem} with respect to
the variables $\vy$, and then take the limits $\eps,\eps'\downarrow
0$. One thus obtains the following

\bigskip

\noindent {\bf Theorem \cite{EL2}:} {\it Let 
\eq
\hat F(\vx;\vn) = \cP(\vn;\vx)\Delta(\vx) \, , 
\quad \vn\in\Z^N 
\eqend
with 
\eq
\Delta(\vx) =  \prod_{1\leq j<k\leq N}\tet(x_k-x_j)^\lambda 
\eqend 
and 
\eq
\cP(\vn;\vx) = 
\lim_{\eps\downarrow 0} 
\int_{-\pi}^{\pi}\f{dy_1}{2\pi}\ee{\ii n_1 y_1}\cdots  
\int_{-\pi}^{\pi}\f{dy_N}{2\pi}\ee{\ii n_N y_N}
\f{ \prod_{1\leq j<k\leq N} \check 
b_{2\eps}(y_{j}-y_k)^{\lambda}}
{\prod_{j,k=1}^N
\check b_{\eps}(y_j-x_k)^{\lambda}} 
\eqend
where 
\eq
 \check b_\eps(r) = 
\left( 1- \ee{\ii r-\eps} \right)
\prod_{m=1}^\infty\left[  \left(1-q^{2m} \ee{\ii r-\eps} 
\right)\left(1-q^{2m} \ee{-\ii r-\eps} \right) \right] \: . 
\eqend
Then the eCS differential operator defined in Eqs.\ \Ref{eCS}, \Ref{V}, 
\Ref{tet} and \Ref{ga} obeys  
\eqa
\label{thm}
H_N \hat F(\vx;\vn) = \cE_0(\vn) \hat F(\vx;\vn) - \ga 
\sum_{1\leq j<k\leq N}
\sum_{n =1}^\infty  \nu  \biggl( 
\f{1}{1-q^{2n}}\hat F(\vx;\vn+ n \vE_{jk}) + \nonu + 
\f{q^{2n}}{1-q^{2n}}\hat F(\vx;\vn - n \vE_{jk}) 
\biggr) 
\eqaend
where
\eq
\cE_0(\vn) = \Bigl[ n_j + \half\lambda(N-1-2j) \Bigr]^2 
\eqend
and $(\vE_{jk})_\ell = \delta_{j\ell}-\delta_{k\ell}$ for 
$j,k,\ell = 1,\ldots, N$. 
}

\bigskip 

\noindent {\em Outline of Proof:} We use Eq.\ \Ref{F} and insert 
$b_\eps(r)= \ee{-\ii r/2} \check b_\eps(r)$ which yields
$$
F^{0,\eps}(\vx;\vy) = const.\ 
\ee{\ii N\lambda \sum_{j=1}^N(x_j-y_j)/2} 
\ee{-\ii\lambda\sum_{1\leq j<k\leq N}(y_j-y_k)/2}
\check \cP^\eps(\vx;\vy) \Delta(\vx)
$$
where 
\eq
\check \cP^\eps(\vx;\vy) = \f{ \prod_{1\leq j<k\leq N} \check 
b_{2\eps}(y_{j}-y_{k})^{\lambda}}
{\prod_{j,k=1}^N
\check b_{\eps}(x_j-y_k)^{\lambda}}  
\eqend
is periodic in the $y_j$.
It is not difficult to show that Eq.\ \Ref{rem1} implies
\eq
\label{abcc}
\biggl[ H_N(\vx) - H_N^{2\eps}(\vy)  \biggr] 
\ee{-\ii N \lambda\sum_{j=1}^N y_j (N+1-2j)/2}
\check \cP^\eps(\vx;\vy) \Delta(\vx) \simeq 0   
\eqend
where we dropped the phase factor describing a center-of-mass 
motion which is the same in $\vx$ and $\vy$ and thus does not 
contribute,\footnote{See Ref.\ \cite{EL2} for the (trivial) details.}
and we used 
$
\sum_{j<k} (y_j-y_k) = \sum_{j} (N+1-2j) y_j . 
$ 
We now can take the Fourier transform of Eq.\ \Ref{abcc}, i.e.,
apply to it $(2\pi)^{-N} \int d^N\vy \, \ee{\ii\vP\cdot\vy }$
where the Fourier variables need to be chosen as
\eq
P_j = n_j  + \half \lambda (N+1-2j)\, , \quad n_j\in\Z
\eqend
so as to compensate the non-periodicity in the $y_j$. 
With that we obtain Eq.\ \Ref{thm}: the term on the l.h.s.\ 
is obvious, the first term on the r.h.s. comes from the derivative 
terms in $H_N^{2\eps}(\vy)$ and partial integration (note that 
$\cE_0 = \sum_j P_j^2$), and the other terms come from 
the potentials $V_{2\eps}(y_j-y_k)$ where we used 
Eq.\ \Ref{V1}. \QED
\bigskip

\begin{Rem} To complete this proof one
needs to show that the correction term $R^{\eps,\eps'}(\vx;\vy)$ in
Eq.\ \Ref{rem} does not contribute, i.e., that for this term the
limits $\eps,\eps'\downarrow 0$ commute with Fourier
transformation. This is plausible, and from our explicit formulas one
can prove this by straightforward computations which, however, is
somewhat tedious. In Ref.\ \cite{EL2} we will give an alternative
proof avoiding this technicality.
\end{Rem} 

This Theorem yields a recursive procedure to construct
eigenfunctions as linear combinations of the functions 
$\hat F(\vx;\vn)$ \cite{EL2}.

\begin{Rem} The functions $\hat F(\vx;\vn)$ in Ref.\ \cite{EL0} differ from
the ones here by the phase factor $\exp(\ii N\lambda\sum_j x_j/2)$
describing a center-of-mass motion. It is easy to see that this
(trivial) change accounts for the different formulas for $\cE_0(\vn)$
(the formula given here agrees with the eigenvalues of the Sutherland
model given in Ref.\ \cite{Su}).

\end{Rem} 

\bigskip

\noindent {\bf Acknowledgements:} 
I thank Alan Carey and Alexios Polychronakos 
for their interest and helpful 
discussions. This work was supported by the Swedish 
Natural Science Research Council (NFR). 



\section*{Appendix~A: Regularized elliptic functions}
In this Appendix we derive various identities related to the functions
defined in Eqs.\ \Ref{beps} and \Ref{expl}, i.e., \eq
C_\eps(r)=\sum_{n=1}^\infty \f{1}{n} \left( \f{1}{1-q^{2n} }\, \ee{\ii
nr-n\eps} + \f{q^{2n} }{1-q^{2n} }\, \ee{-\ii n r-n\eps} \right)
\eqend and $b_\eps(r)= \ee{-\ii r/2 - C_\eps(r) }$.  In particular we
prove Eq.\ \Ref{b1} and derive three different representations for the
function \eq V_\eps(r) =- \f{\partial^2}{\partial r^2}\log b_\eps(r) =
\f{\partial^2}{\partial r^2}C_{\eps}(r) \eqend which also prove that
it is a regularization of the interaction potential defined in Eq.\
\Ref{V}, $V_0(r)=V(r)$. We also make precise the relation of $V(r)$ to
Weierstrass' $\wp$ function.

From the formulas given above we get immediately our first representation, 
\eq
\label{V1}
V_\eps(r) = -\sum_{n=1}^\infty n\left[ \f{1}{1-q^{2n}}\ee{\ii n r} + 
 \f{q^{2n}}{1-q^{2n}} \ee{-\ii n r} \right] \ee{-n\eps} \: . 
\eqend
By expanding $1/(1-q^{2n})$ in geometric series we obtain  
\nonueqa
C_\eps(r)=\sum_{n=1}^\infty \f{1}{n}\left[ \ee{\ii nr-n\eps} + 
\sum_{m=1}^\infty q^{2nm}\left( \ee{\ii nr-n\eps} + \ee{-\ii n r-n\eps} 
\right) \right]\, , 
\nonueqaend
and interchanging the summation we can do the $n$-sum and obtain
\nonueqa 
C_\eps(r)= - \log\left(1- \ee{\ii r-\eps} \right) - 
\sum_{m=1}^\infty \left[ \log\left(1-q^{2m} \ee{\ii r-\eps} \right) + 
 \log\left(1-q^{2m} \ee{-\ii r-\eps} \right) \right] \, , 
\nonueqaend
i.e., 
\eq
\label{C2}
C_\eps(r) = 
-\log\left( \ee{\ii r/2} b_{\eps}(r)\right) 
\eqend
with 
\nonueqa
b_\eps(r) = \ee{-\ii r/2}\left( 1- \ee{\ii r-\eps} \right)
\prod_{m=1}^\infty\left[  \left(1-q^{2m} \ee{\ii r-\eps} 
\right)\left(1-q^{2m} \ee{-\ii r-\eps} \right) \right]
\nonueqaend
identical with $b_\eps(r)$ in Eq.\ \Ref{b1}. This
also yields our second representation Eq.\ \Ref{V2}, 
which proves that $V_0(r)$ is identical with $V(r)$ defined in Eq.\ \Ref{V}.

Inserting $q=\exp{(-\beta/2)}$ we can also write
\nonueqa
C_\eps(r) =  
-\sum_{m=0}^\infty  \log\left(1- \ee{\ii (r + \ii \beta m +\ii \eps) } \right) - 
 \sum_{m=1}^\infty \log\left(1- \ee{-\ii ( r - \ii \beta m- \ii \eps) } \right)  \: , 
\nonueqaend 
and using 
\nonueqa
\f{\partial^2}{\partial r^2} \log\left(1- \ee{\pm \ii (r \pm \ii \beta m\pm \ii\eps) } \right) 
= 
\f{\partial^2}{\partial r^2} \log\left( \sin\half(r \pm \ii \beta m \pm \ii \eps ) \right) \\
= -\f{1}{4\sin^2\half(r \pm \ii \beta m \pm \ii \eps ) } 
\nonueqaend
we obtain 
\eq
\label{V3}
V_\eps(r) = 
\sum_{m=0}^\infty  \f{1}{4\sin^2\half(r + \ii \beta m + \ii\eps ) } + 
\sum_{m=1}^\infty \f{1}{4\sin^2\half(r - \ii \beta m - \ii\eps ) }  \: . 
\eqend 
From this representation it is obvious that the function 
$V(z)\equiv V_0(z)$, $z\in\C$,  is doubly periodic
with periods $2 \pi$ and $\ii\beta$, it has a single pole of order 2
in each period-parallelogram, 
$V(z) - z^{-2}$ is analytic in some neighborhood
of $z=0$ and equal to 
$1/12-\sum_{m=1}^\infty (1/2) \sinh^{-2}( \beta m/2 )$
in $z=0$. These facts imply (see e.g.\ \cite{Bateman}, Sect.\ 13.12)
\eq
\label{wp}
V(r) = \wp(r| \pi, \half \ii\beta) + \f{1}{12} - 
\sum_{m=1}^\infty\f{1}{2\sinh^2\half ( \beta m ) }  \: . 
\eqend

\section*{Appendix~B: On the loop group of U(1)}
In this Appendix we clarify the relation of our construction
of the quantum field theory model of anyons and the representation
theory of loop groups. We also explain and prove the physical 
interpretation of the representations which we are using. 

\subsection*{B.1 Loop group of U(1)} 
Let $\cG=$Map($S^1$,U(1)) be the set of all $C^{1/2}$ 
functions $S^1\to \U(1)$, i.e.,
it contains all functions of the form $\ee{\ii f}$ with
\eq
f(x) = w x +\sum_{n\in\Z}\alpha_n\ee{\ii nx}\: , 
\eqend
$-\pi\leq x\leq \pi$, 
$w$ an integer called {\it winding number},  
and complex $\alpha_n$ such that 
\eq
\label{loopcond1}
\alpha_n=\overline{\alpha_{-n}}\, , \quad 
\sum_{n=1}^\infty n|\alpha_n|^2<\infty \: . 
\eqend
This set has an obvious group structure, i.e., the product is 
by point-wise multiplication and the inverse 
by point-wise complex conjugation. 
The group  $\cG$ has an interesting central extension 
$\hat\cG\, :=\,  \U(1)\times \cG  = 
\{ (c,\ee{\ii f}) |c\in \U(1), \ee{\ii f}\in\cG \}$
with the group structure given by $(c,\ee{\ii f})^{-1}\, := \, 
(\bar c,\ee{-\ii f} ) $ and  
\eqa 
\label{def}
 (c,\ee{\ii f}) (c',\ee{\ii f'}) \, : = \,  
(c c' \ee{-\ii S(f,f')/2 } , \ee{\ii(f+f') } )\: , 
\eqaend
where
\eqa
S(f,f') \, &:=& \, 
  w \alpha'_0 -w' \alpha_0 
- \ii \sum_{n\in\Z} n \alpha_{-n}\alpha'_n 
\eqaend
is a 2-cocycle of the group $\cG$, i.e., it obeys relations such that Eq.\
\Ref{def} indeed defines a group structure on $\hat\cG$. Note that
the conditions in Eq.\ \Ref{loopcond1} and Cauchy's inequality ensure
that $S(f,f')$ is always well-defined.

\subsection*{B.2 Representations of the loop group of U(1)} 
Given a representation of the algebra $\cA_0$ defined in Eqs.\ 
\Ref{12} and Eq.\ \Ref{un}, one has, at least formally, also
a unitary representation of the group $\hat \cG$. This can be seen as 
follows: Defining 
\eq
\Gamma(\ee{\ii f})\, := \, \ee{\ii \alpha_0 Q/2} R^{w} \ee{\ii \alpha_0 Q/2} 
\ee{\ii J(\alpha) }
\eqend
with 
\eq
J(\alpha) \, := \, \sum_{n\in\Zp} \alpha_n\b(-n) \: , 
\eqend
a simple computation using the Hausdorff formula Eq.\ \Ref{Hausdorff} 
together with 
$$
\ccr{J(\alpha)}{J(\alpha')} = \sum_{n\in\Z} n\alpha_{-n} \alpha'_n \id 
$$
(which follows from Eq.\ \Ref{12}) and Eq.\ \Ref{QRQ} yields 
\eq
\Gamma(\ee{\ii f})\Gamma(\ee{\ii f'}) = \ee{-\ii S(f,f')/2}\Gamma(\ee{\ii (f+f')})\: .
\eqend
Moreover, it is easy to see that $\Gamma(\ee{\ii f})^* = \Gamma(\ee{-\ii f})$. 
Thus $(c,\ee{\ii f})\to c\Gamma(\ee{\ii f})$ is a 
unitary representation of $\hat \cG$. 

\newcommand{\mm}{\underline{m}}

\subsection*{B.3 Zero- and finite temperature representations} 
The standard representation $\pi_0$ of the algebra $\cA_0$ is on a
Hilbert space $\cF_0$ and completely characterized by the following 
conditions, 
\eq
\b_0(n)\Omega_0=0\quad \forall n\geq 0\, ,\quad 
\pr{\Omega_0}{R_0^w\Omega_0}=\delta_{w,0} 
\quad \forall w\in\Z
\eqend
where we write $\b_0(n)\equiv\pi_0(\b(n))$ and similarly for $R$ and $Q$, 
and $\Omega_0\in\cF_0$ is the highest weight state. 
These conditions imply that the states
\eq
\label{eta1}
\eta_w(\mm) = \prod_{n=1}^\infty 
\f{\b_0(-n)^{m_{n}}}{\sqrt{m_{n}!n^{m_{n}}}} \, 
R_0^w \, \Omega\: , \quad 
m_{n}\in\N_0 \: , \sum_{n=1}^\infty 
m_{n}<\infty \: , w \in\Z
\eqend
provide a complete orthonormal basis in $\cF_0$, and it is
easy to check that these states all are eigenstates of the
Hamiltonian 
\eq
\label{Ham0}
H_0 =\f{a}{2}Q_0^2 + \sum_{n=1}^\infty \b_0(-n)\b_0(n) 
\eqend
with corresponding eigenvalues
\eq
 \cE_w(\mm) = \f{a}{2}w^2 + \sum_{n=1}^\infty m_n n \: . 
\eqend
This also shows that $H_0$ defines a self-adjoint operator. 
Moreover, for $a>0$ the Hamiltonian $H_0$
is non-negative and $\Omega_0$ is the ground-state, 
i.e., the unique eigenstate of $H_0$ with minimum eigenvalue. 
The representation $\pi_0$ is therefore interpreted as 
{\em zero temperature representation}, and for elements $A$ in $\cA_0$, 
$<A>_\infty\, := \, \pr{\Omega_0}{A_0 \Omega_0}$ is interpreted as the 
ground state expectation value of $A$ ($A_0\equiv \pi_0(A)$). 
More generally, the finite temperature expectation value 
is defined as
\eq
\label{finiteT}
<A>_\beta \, : = \, \f{1}{\cZ}\, \Tr_{\cF_0}(\ee{-\beta H_0} A_0) \, , 
\quad   \cZ\, := \Tr_{\cF_0}(\ee{-\beta H_0})
\eqend
with a normalization constant $\cZ$ called {\em partition function}
and $ \Tr_{\cF_0}$ the Hilbert space trace in $\cF_0$. The parameter
$\beta>0$ is interpreted as the inverse temperature. 

The representation discussed in the main text are on the Hilbert space
$\cF=\cF_0\otimes \cF_0$ with $\Omega=\Omega_0\otimes\Omega_0$ and
\eq
\b_1(n) = \b_0(n)\otimes \id\, , \quad \b_2(n) = \id \otimes \b_0(n)  
\eqend
and similarly for $R_{1,2}$. We now can formulate 
the precise meaning of the statement in Remark \ref{2.2} in 
the main text: 

\begin{Prop}\label{PB}
Let $\cA_0'$ be the dense set in $\cA_0$ containing all finite
linear combinations of elements 
\eq
\label{A}
A = R^u Q^v \prod_{n=1}^\infty \b(-n)^{k_n}\b(n)^{\ell_n}
\eqend
where $u\in\Z$ and $v,k_n,\ell_n\in\N_0$ and such that
$\sum_{n=1}^\infty (k_n+\ell_n)<\infty$.
For the representation
$\pi$ defined in Eqs.\ \Ref{bn} and \Ref{expl} and for all $A\in\cA_0'$, 
the following holds true
\eq
\label{ft}
\pr{\Omega}{ \pi(A) \Omega}_{\cF} \, = \, <A>_\beta
\eqend
if $q=\exp{(-\beta/2)}$ and $a\to \infty$.
\end{Prop}

\noindent {\em Proof:} It is obviously enough to prove Eq.\ \Ref{ft}
for all $A$ in Eq.\ \Ref{A}.  We do this by explicit computations. We
first compute
\nonueqa
 \Tr_{\cF_0}( \ee{-\beta H_0} A_0 ) = 
\sum_{w\in\Z}\; \sum_{m_1,m_2,\ldots =0}^\infty 
\pr{\eta_w(\mm)}{\ee{-\beta H_0} 
A_0 \eta_w(\mm)} \nonu
= \sum_{w\in\Z} \; \sum_{m_1,m_2,\ldots =0}^\infty 
\ee{-\beta\cE_w(\mm)} \delta_{u,0} 
w^v\prod_{n=1}^\infty \delta_{k_n,\ell_n}
\f{(m_n+k_n)! n^{m_n+k_n}}{m_n! n^{m_n}}
\nonu 
= \delta_{u,0} \sum_{w\in\Z}\ee{-\beta a w^2/2} w^v 
\prod_{n=1}^\infty  \delta_{k_n,\ell_n} n^{k_n} 
\sum_{m_n=0}^\infty \f{(m_n+k_n)!}{m_n!} \ee{-\beta m_n n} \: ,  
\nonueqaend
and with
$$
\sum_{m=0}^\infty  \f{(m+k)!}{m!} x^{m} = \sum_{m=0}^\infty \f{d^k}{dx^k}
x^{m+k} 
= \f{d^k}{dx^k} \f{1}{1-x} = k! \f{1}{(1-x)^{k+1}}
$$
for $x=\ee{-\beta n}$, we get
\nonueqa
 \Tr_{\cF_0}(\ee{-\beta H_0} A_0)
=  \delta_{u,0} \sum_{w\in\Z}\ee{-\beta a w^2/2} w^v 
\prod_{n=1}^\infty  \delta_{k_n,\ell_n} n^{k_n} k_n! 
\f{1}{(1-\ee{-\beta n})^{k_n+1}} \: . 
\nonueqaend
In particular,
\eq
\cZ =  \sum_{w\in\Z}\ee{-\beta a w^2/2}
\prod_{n=1}^\infty\f{1}{(1-\ee{-\beta n})}
\eqend
and
\eq
\label{rhs}
<A>_\beta \, = \,  \delta_{u,0}
\f{ \sum_{w\in\Z}\ee{-\beta a w^2/2} w^v }{\sum_{w\in\Z}\ee{-\beta a w^2/2}}
\prod_{n=1}^\infty  \delta_{k_n,\ell_n} n^{k_n} k_n! 
\f{1}{(1-\ee{-\beta n})^{k_n}} \: . 
\eqend

We now compute the l.h.s.\ of Eq.\ \Ref{ft} using Eqs.\ \Ref{bn} and 
\Ref{ccr}--\Ref{ROm},
\nonueqa
\pr{\Omega}{ \pi(A) \Omega}_{\cF} = 
\biggl<\Omega, R_1^u Q_1^v \prod_{n=1}^\infty (c_{-n}\b_1(-n) + s_{-n}\b_2(n))^{k_n}\nonu
\times (c_{n}\b_1(n) + s_n\b_2(-n))^{\ell_n}\Omega\biggl>_{\cF}
= \delta_{u,0}\delta_{v,0} \prod_{n=1}^\infty \delta_{k_n,\ell_n}
|c_n|^{k_n} n^{k_n} k_n! \: , 
\nonueqaend
and with Eq.\ \Ref{expl},
\eq
\pr{\Omega}{ \pi(A) \Omega}_{\cF} = \delta_{u,0}\delta_{v,0} \prod_{n=1}^\infty 
\delta_{k_n,\ell_n}n^{k_n} k_n! \left( \f{1}{1-q^{2n}} \right)^{k_n} \: . 
\eqend
This obviously coincides with Eq.\ \Ref{rhs} for $q=\exp{(-\beta/2)}$
and $a\to\infty$. 
\QED

\begin{Rem}\label{A1}
From a physics point of view, the more 
natural Hamiltonian for the anyon system would be $H_0$ as in 
Eq.\ \Ref{Ham0} with $a=1/2$ (and not $a\to\infty$). It is 
therefore interesting to note that the finite temperature 
state for this Hamiltonian and arbitrary $a>0$ is
\eq
\Omega' = \f{1}{z^{1/2}}
\sum_{w\in\Z} (R_1 R_2)^w \ee{-a w^2/2} \Omega\, , \quad 
z\, := \sum_{w\in\Z}  \ee{-a w^2/2} 
\eqend
which is easily proven by generalizing our argument above.
It is also easy to compute the anyon correlation functions 
with that state. However, we find that this state is useful 
for the eCS system only if we choose $a\to\infty$. 

\end{Rem}

\section*{Appendix~C: Proofs}

\subsection*{C.1 Proof of Lemma~\ref{L1}}
We will prove this Lemma by direct computation. For that we will
need several identities which we derive first. We will need
\eq
\label{Wt1}
\ccr{Q}{\phi_\eps(x)} = \phi_\eps(x) 
\eqend
which follows trivially from
the relations in Eq.\ \Ref{12} and the definition of 
$\phi_\eps(x)\equiv\phi^1_\eps(x)$ in  Eqs.\ \Ref{K}--\Ref{phi}.
It is useful to 
write $\bb_\eps(x) = \nu Q+ \rho^+_\eps(x) + \rho^-_\eps(x)$
with the creation- and annihilation parts 
\eq
\label{bbpm}
\rho_\eps^\pm(x) \, := \, \sum_{n=1}^\infty
\left( c_{\pm n}\, \ee{\mp \ii nx-n\eps} \b_1(\mp n) + 
s_{\mp n} \, \ee{\pm \ii nx-n\eps} \b_2(\mp n) \right) \: . 
\eqend
We note that $\rho_\eps^\pm(x) = \partial K_\eps^\pm(x)/\partial x$, and 
it follows therefore from Eq.\ \Ref{beps} that 
\eq
\label{1st}
\f{\partial}{\partial x}\phi_\eps(x) = \, -\ii \xxa 
\left[ \nu^2 Q + \nu\rho^+_\eps(x) + \nu\rho^-_\eps(x) \right] \phi_\eps(x)\xxe
\, \equiv  \, -\ii \nu \xxa \bb_\eps(x) \phi_\eps(x)\xxe
\eqend
which implies 
\eqa
\label{2nd}
\f{\partial^2}{\partial x^2}  \phi_\eps(x) =  
-\ii \nu \f{\partial}{\partial x}\xxa \bb_{\eps}(x) \phi_\eps(x) \xxe
\, = -\ii \nu\xxa \bb'_{\eps}(x) \phi_\eps(x) \xxe - 
\nu^2 \xxa \bb_{\eps}(x)^2 \phi_\eps(x) \xxe  \: . 
\eqaend
Moreover, a simple computation using Eqs.\ \Ref{Kpm}, \Ref{ccr}, \Ref{cnsn1} 
and  \Ref{cnsn2} yields  
$$
\ccr{\rho^\mp(y)_{\eps'}}{K^\pm_\eps(x)}
= \sum_{n=1}^\infty 
\ii \left[  (1+|s_n|^2)\,   \ee{\mp \ii nr -n\eps}
- |s_n|^2 \ee{\pm \ii n r-n\eps} \right] \equiv 
2\pi \ii \delta_{\teps}^\mp (r)  \pm \jj_{\teps}(r)
$$
where
\eq
\delta^\pm_\eps(r) \, := \,  \f{1}{2\pi} \sum_{n=1}^\infty \ee{\pm\ii nr-n\eps} 
\eqend
and 
\eq
\label{jj}
\jj_\eps(r) \, := \, 2 \sum_{n=1}^\infty  |s_n|^2 \sin(nr) \ee{-n\eps} \: . 
\eqend
Using that we compute ($t\in\C$) 
\nonueqa
\xxa\ee{t \bb_{\eps'}(y)} \xxe
\phi_\eps(x)= 
\ee{t\nu Q} \ee{t \bb^+_{\eps'}(y)}\ee{t \bb^-_{\eps'}(y)}
\ee{-\ii \nu^2 Qx/2}\, R \, \ee{- \ii \nu^2 Qx/2} 
\ee{ - \ii\nu K^+_\eps(x) } \ee{ - \ii\nu K^-_\eps(x) } 
\\
= \ee{t\nu/2 } \ee{-\ii t\nu \ccr{\bb^-_{\eps'}(y) }{K^+_\eps(x)} }
\ee{(-\ii \nu^2 x + \nu t)Q/2}\, R\,  \ee{(-\ii \nu^2 x + \nu t)Q/2} 
 \ee{t \bb^+_{\eps'}(y)}\ee{ - \ii\nu K^+_\eps(x) }\ee{t \bb^-_{\eps'}(y)} \ee{ - \ii\nu K^-_\eps(x) } \\
\equiv \ee{ t\nu \Delta^-_{\eps+\eps'}(x-y)}\xxa \ee{t \bb_{\eps'}(y)} \phi_\eps(x)\xxe
\nonueqaend
where 
$\Delta^-_{\eps}(r) =  1/2 + 2\pi\delta^-_{\eps}(r) - \ii j_{\eps}(r) $. 
Similarly, 
\nonueqa
\phi_\eps(x)\xxa\ee{t \bb_{\eps'}(y)} \xxe\, = \, 
\ee{-t\nu/2 }\ee{-\ii t\nu\ccr{ K^-_\eps(x)}{ \bb^+_{\eps'}(y)}  }
\xxa \ee{t \bb_{\eps'}(y)} \phi_\eps(x)\xxe \nonu 
\equiv 
\ee{-t\nu \Delta^+_{\eps+\eps'}(x-y)}\xxa \ee{t \bb_{\eps'}(y)} \phi_\eps(x)\xxe 
\nonueqaend
where $\Delta^+_{\eps}(r) =  1/2 + 2\pi\delta^+_{\eps}(r) + \ii j_{\eps}(r)$. 
Thus 
\eq
\label{Wtall}
\ccr{\xxa \ee{t \bb_{\eps'}(y) } \xxe}{\phi_\eps(x) }
= \left( \ee{ t \nu \Delta_{\teps}^-(r)} 
- \ee{ - t\nu \Delta_{\teps}^+(r)} 
\right) \xxa \ee{t \bb_{\eps'}(y) }  \phi_\eps(x)  
\xxe\: 
\eqend
with $\teps=\eps+\eps'$, $r=x-y$, and  
\eq
\label{chi}
\Delta^\pm_{\eps}(r) =  \f{1}{2} + 2\pi \delta_\eps^\pm(r) 
\pm \ii \jj_{\eps}(r) \: .  
\eqend
We will also need the following three identities,   
\eqa
\label{Del3}
\Delta_{\eps}^-(r) + \Delta_{\eps}^+(r) &=& 2\pi\delta_{\eps}(r)\nonu
\Delta_{\eps}^-(r)^2 - \Delta_{\eps}^+(r)^2 &=& 
2\pi \ii\delta'_\eps(r) - 4\pi\ii \jj_\eps(r)\delta_\eps(r) \\
\Delta_{\eps}^-(r)^3 + \Delta_{\eps}^+(r)^3 &=& 
-\pi\delta_\eps''(r) + \f{1}{2}\pi\delta_\eps(r) + 6\pi\jj_\eps(r)i\delta_\eps'(r)
-6\pi\jj_\eps(r)^2\delta_\eps(r)  \nonumber
\eqaend
where 
$\delta_\eps(r)$ is the regularized $\delta$-function defined in Eq.\ \Ref{delta} 
and the prime indicates differentiation with respect to the argument $r$.
The first of these identities is obvious, and the second and third identities can be 
proven by straightforward computations using 
$$
\f{1}{2} + 2\pi \delta_\eps^\pm(r) = \f{1+\ee{\pm \ii r-\eps}}{2( 1-\ee{\pm \ii r-\eps} )} 
= \pm  \f{\ii}{2}\cot\half(r\pm\ii\eps)
$$
and
$$
\cot^2(\half z) = -2\f{d}{dz} \cot(\half z) -1\, ,  \quad  
\cot^3(\half z) = 2\f{d^2}{dz^2}\cot(\half z)  - \cot(\half z) \: . 
$$
For the convenience of the reader we give some details of these computations,
\nonueqa
\Delta_{\eps}^\pm (r)^2 
= (\pm\ii)^2\left[\f{1}{2}\cot\half(r \pm  \ii\eps) + \jj_\eps(r) \right]^2  = \nonu
-\f{1}{4} \cot^2\half(r \pm \ii\eps) - \jj_\eps(r) \cot\half(r \pm \ii\eps) - \jj_\eps(r)^2 = \nonu
= \left(  \f{1}{2}\f{\partial}{\partial r} -  \jj_\eps(r) \right) \cot\half(r \pm \ii\eps)  
+ \f{1}{4} -  \jj_\eps(r)^2 
\nonueqaend
which together with 
$$
\cot\half(r - \ii\eps) -
\cot\half(r + \ii\eps) = 4\pi\ii \delta_\eps(r)
$$
proves the second identity in \Ref{Del3}, and similarly
\nonueqa
\pm \ii  \Delta_{\teps}^\pm (r)^3 
= (\pm\ii)^4 \left[\f{1}{2}\cot\half(r \pm  \ii\eps) + \jj_\eps(r) \right]^3  = \nonu
\f{1}{8} \cot^3\half(r \pm \ii\eps) + \f{3}{4} \jj_\eps(r) \cot^2\half(r \pm \ii\eps) + 
 \f{3 }{2} \jj_\eps(r)^2 \cot\half(r \pm \ii\eps) + \jj_\eps(r)^3 = \nonu
= \left( \f{1}{4}\f{\partial^2}{\partial r^2} -\f{1}{8} - 
\f{3}{2}\jj_\eps(r)\f{\partial}{\partial r}
+  \f{3 }{2} \jj_\eps(r)^2 
 \right) \cot\half(r \pm \ii\eps)  
- \f{3}{4} \jj_\eps(r) -  \jj_\eps(r)^3  
\nonueqaend
implies the third identity. 

Differentiating Eq.\ \Ref{Wtall} three times with respect to $t$ and setting $t=0$
yields
\nonueqa
\ccr{\xxa \bb_{\eps'}(y)^3 \xxe}{\phi_\eps(x) }
= \nu^3 \left[ \Delta_{\teps}^-(r)^3 + \Delta_{\teps}^+(r)^3 
\right] \phi_\eps(x) 
+ 3\nu^2  \left[ \Delta_{\teps}^-(r)^2 - \Delta_{\teps}^+(r)^2 
\right] 
\nonu\times 
\xxa \bb_{\eps'}(y) \phi_\eps(x) \xxe +  
3\nu  \left[ \Delta_{\teps}^-(r) + \Delta_{\teps}^+(r) 
\right]  \xxa \bb_{\eps'}(y)^2   \phi_\eps(x) \xxe  \: ,  
\nonueqaend
and by inserting the identities in Eq.\ \Ref{Del3} we obtain
\nonueqa
\ccr{\f{1}{6\pi}\xxa \bb_{\eps'}(y)^3 \xxe}{\phi_\eps(x) }
= \nu^3 \left[ 
-\ff{1}{6}\delta_{\teps}''(r) + \f{1}{12}\delta_{\teps}(r) + \jj_{\teps}(r) \delta'_{\teps}(r)
- \jj_{\teps}(r)^2\delta_{\teps}(r)  
\right] \phi_\eps(x) + \nonu
+ \nu^2  \left[\ii\delta'_{\teps}(r) - 2\ii j_{\teps}(r)\delta_{\teps}(r) 
\right] \xxa \bb_{\eps'}(y) \phi_\eps(x) \xxe +  
\nu  \delta_{\teps}(r) \xxa \bb_{\eps'}(y)^2   \phi_\eps(x) \xxe  \: .  
\nonueqaend
With that we compute
\nonueqa
\ccr{\tW^3}{\phi_\eps(x)} = \lim_{\eps'\downarrow 0} 
\int_{-\pi}^{\pi} \ccr{\f{1}{6\pi}\xxa \bb_{\eps'}(y)^3 
\xxe}{\phi_\eps(x) }-\nu^3 c_\eps\ccr{Q}{\phi_\eps(x)}
\nonu
= 
\nu^3 a_\eps\phi_\eps(x) + 
\ii \nu^2 \xxa \bb'_{\eps}(x) \phi_\eps(x) \xxe  + 
\nu  \xxa \bb_{\eps}(x)^2   \phi_\eps(x) \xxe +  
\xxa rest_{\eps}(x)   \phi_\eps(x) \xxe
\nonueqaend
with 
$$
a_\eps\, := \,  \f{1}{12} - \jj'_{2\eps}(0)  -  
\lim_{\eps'\downarrow 0} \int_{-\pi}^{\pi} dy\, \delta_{\teps}(r) \jj_{\teps}(r)^2  -c_\eps 
$$
where we used Eq.\ \Ref{Wt1} and the following relations,  
\nonueqa
\int_{-\pi}^{\pi} dy\, \delta_{\teps}''(x-y) = 0  
\nonu
\int_{-\pi}^{\pi} dy\, \delta_{\teps}(x-y) = 1  
\nonu
\int_{-\pi}^{\pi} dy\, \jj_{\teps}(x-y) \delta'_{\teps}(x-y) = -j'_{2\eps+2\eps'}(0)  
\nonu
\int_{-\pi}^{\pi} dy\,  \delta'_{\teps}(x-y) \bb_{\eps'}(y) = \bb'_{\eps+2\eps'}(x) 
\nonueqaend
(note the sign in the last relation!) 
which all are simple consequences of the definitions, and a rest term 
\eq
rest_\eps(x) \, := \, 
 \lim_{\eps'\downarrow 0} \int_{-\pi}^{\pi} dy\, 
\delta_{\teps}(x-y) [ - 2\ii\nu^2 \jj_{\teps}(x-y) \bb_{\eps'}(y) + \nu\bb_{\eps'}(y)^2 ]  - 
\nu\bb_{\eps}(x)^2   
\eqend
which we expect to vanish in the limit $\eps\downarrow 0$ (recall that $\jj_\eps(0)=0$). 
For the following computations we find it convenient to write (cf.\ Eq.\ \Ref{jj})
\eq
\label{jj1}
\jj_\eps(r) =  \sum_{n\in\Z}  \hj(n)\, \ee{\ii nx-|n|\eps}\: , 
\quad \hj(|n|) = -\ii|s_n^2|= -\hj(-|n|)   \: . 
\eqend
With that, 
\nonueqa
\lim_{\eps'\downarrow 0} \int_{-\pi}^{\pi} dy\, \delta_{\teps}(r) \jj_{\teps}(r)^2 = 
\lim_{\eps'\downarrow 0} \sum_{n_1,n_2,n_3\in\Z}\f{1}{2\pi}
\int_{-\pi}^{\pi} dr\,  \ee{\ii n_1 r- |n_1|\teps } 
 \hj(n_2)\ee{\ii n_2 r-|n_2|\teps} \hj(n_3)\ee{\ii n_3 r-|n_3|\teps} \\ 
= \sum_{m,n\in\Z} \hj(m)\hj(n)\ee{-(|m|+|n|+|m+n|)\eps} =  
\sum_{m,n=1}^\infty 2 |s_m|^2|s_n|^2\ee{-(m+n)\eps}\left(\ee{-|m-n|\eps} - 
\ee{-|m+n|\eps} \right) \: , 
\nonueqaend
and with $j'_{2\eps}(0) = 2\sum_{n=1}^\infty n|s_n|^2 \ee{-2|n|\eps}$
(cf.\ \Ref{jj}) and Eq.\ \Ref{cceps} we get $a_\eps\equiv 0$. 
Using then Eq.\ \Ref{2nd} to eliminate the term with 
$\bb_{\eps}(x)^2$ we obtain  
\nonueqa
\ccr{\tW^3}{\phi_\eps(x)} \, &=& \, 
- \f{1}{\nu}\f{\partial^2}{\partial x^2} \phi_\eps(x)      
+  \ii( \nu^2 - 1) \xxa 
\bb'_\eps(x) \phi_\eps(x)\xxe + \xxa rest_\eps \phi_\eps(x)\xxe 
\nonueqaend
which coincides with Eq.\ \Ref{Wt3} provided that
\eq
\label{final}
rest_\eps(x) = \eps \rr_\eps(x)
\eqend
with $\rr_\eps(x)$ defined in Eq.\ \Ref{rr}. We thus are left to prove Eq.\ 
\Ref{final}. For that we compute the rest term explicitly. Firstly, recalling 
Eqs.\ \Ref{jj1} and \Ref{bb} and writing $\bb(0)=\nu Q$ and $\bb(n)=\b(n)$ 
for $n\neq 0$, 
\nonueqa
(*)_1\, :=\,  \lim_{\eps'\downarrow 0} \int_{-\pi}^{\pi} dy\, 
\delta_{\teps}(x-y) \jj_{\teps}(x-y) \bb_{\eps'}(y) \nonu = 
 \lim_{\eps'\downarrow 0} \sum_{n_1,n_2,n_3\in\Z}\f{1}{2\pi}
\int_{-\pi}^{\pi} dy\,  \ee{\ii n_1(x-y)- |n_1| (\eps+\eps') } 
\hj(n_2)\, \ee{\ii n_2(x-y) -|n_2|(\eps+\eps')}\bb(n_3)\, \ee{\ii n_3 y-|n_3|\eps'}\nonu
= \sum_{m,n\in\Z} \hj(m) \bb(n)\ee{\ii n x}\ee{-(|m|+|m+n|)\eps}\nonu = 
\sum_{m,n=1}^\infty \ii |s_m|^2 
\left( \b(n)\ee{\ii nx} -  \b(-n)\ee{-\ii nx} \right) 
\left(\ee{-|m-n|\eps} - \ee{-|m+n|\eps} \right)\ee{-|m|\eps} \: .   
\nonueqaend
Secondly, 
\nonueqa
(*)_2 \, := \, \lim_{\eps'\downarrow 0}
\int_{-\pi}^{\pi} dy\, 
\delta_{\eps+\eps'}(x-y) \bb_{\eps'}(y)^2    
\nonu
= \lim_{\eps'\downarrow 0} \sum_{n_1,n_2,n_3\in\Z} 
\f{1}{2\pi}
\int_{-\pi}^{\pi} dy\, \ee{\ii n_1(x-y)- |n_1|(\eps+\eps') } 
\bb(n_2)\, \ee{\ii n_2 y -\eps'|n_2| } \bb(n_3)\, \ee{\ii n_3 y -\eps' |n_3|}
\nonu
= \sum_{m,n\in\Z} \bb(m)\bb(n) \ee{\ii (m+n) x -\eps|m+n|} \: ,  
\nonueqaend
and recalling 
$
\bb_\eps(x)^2 =  \sum_{m,n\in\Z} \bb(m)\bb(n) \ee{\ii (m+n) x -\eps (|m|+|n|)} 
$
we obtain
\nonueqa
\xxa [(*)_2 - \bb_\eps(x)^2 ] \xxe  = \sum_{m,n\in\Z} \xxa \bb(m)\bb(n)\xxe \ee{\ii (m+n) x} 
\left(\ee{-\eps(|m+n|) } -  \ee{-(|m|+|n|)\eps} \right)\nonu =  
2\sum_{m,n=1}^\infty \xxa \b(-m)\b(n) \xxe \ee{-\ii(m-n)x}
\left( \ee{-|m+n|\eps}  - \ee{- |m-n|\eps} \right) \: . 
\nonueqaend
We therefore obtain (cf.\ Eq.\ \Ref{rr})  
$$
rest_\eps(x) = -2\ii \nu^2 (*)_1 + \nu \xxa[(*)_2 - \bb_\eps(x)^2 ] \xxe  
\, \equiv \eps \rr_\eps(r) 
$$
which completes the proof of Lemma~\ref{L1}. \QED

\subsection*{C.2 Proof of Lemma~\ref{L2}}
We write $J(\al) =\alpha_0 Q + A^- + A^+$
where $A^\pm = J^\pm(\al)$
and recall from the definition of normal ordering that 
$$
\Phi = \ee{\ii\alpha_0 Q/2}\, R^w\ee{\ii\alpha_0 Q/2} \ee{\ii A^+}\ee{\ii A^-}
$$ and 
\nonueqa
\xxa  \Phi \cC \xxe \, = \sum_{n=1}^\infty n \left[ \b_1(-n)\Phi\b_1(n) +\b_2(-n)\Phi\b_2(n) \right] \: . 
\nonueqaend
To compute the l.h.s.\ of Eq.\ \Ref{PHI1} we use
\nonueqa
\ccr{\b_1(\pm n)}{A^{\pm}} &=& \pm n c_{\mp n} \alpha_{\pm n}\id \\
\ccr{\b_2(\pm n)}{A^{\pm}} &=& \pm n s_{\pm n} \alpha_{\mp n}\id  \quad \forall n\geq 0 
\nonueqaend
following from Eqs.\ \Ref{bn} and \Ref{ccr}. Moreover, we get for $n>0$,
\eqa
\label{b121}
\b_1(n) \Phi = \left. \f{\partial}{\partial t}\,  \ee{t\b_1(n)} \Phi\right|_{t=0} 
=  \left. \f{\partial}{\partial t}\, \ee{\ii t\ccr{\b_1(n)}{A^+} } \Phi \ee{t\b_1(n)}\right|_{t=0} \nonu 
= \Phi \b_1(n) + \ii \ccr{\b_1(n)}{A^+} =  \Phi \b_1(n) +\ii n c_{-n}\alpha_n\id  \: ,  
\eqaend 
and similarly,
\eqa
\label{b122}
\Phi\b_1(-n) = \b_1(-n)\Phi -  \ii \ccr{\b_1(-n)}{A^-} &=& \b_1(-n)\Phi  + \ii n c_{n}\alpha_{-n}\id \nonu
\b_2(n)\Phi = \Phi \b_2(n)\ii +  \ii \ccr{\b_2(n)}{A^+} &=&\Phi \b_2(n)  + \ii n s_{n}\alpha_{-n}\id \nonu
\Phi\b_2(-n) = \b_2(-n)\Phi -  \ii \ccr{\b_2(-n)}{A^-} &=& \b_2(-n)\Phi  + \ii n s_{-n}\alpha_{n}\id \: . 
\eqaend
With that we obtain 
\nonueqa
\cC \Phi +  \Phi\cC -2\xxa  \Phi \cC \xxe \, = \sum_{n=1}^\infty n\xxa \left[ 
\ii n c_{-n}\alpha_n  \b_1(-n) + \ii n c_{n}\alpha_{-n} \b_1(n) \right.\nonu\left. +  
\ii n s_{n}\alpha_{-n}  \b_2(-n) + \ii n s_{-n}\alpha_{n} \b_2(n)   
\right]\Phi \xxe  \, = \,  \xxa \sum_{n\in\Z} \ii n^2 \alpha_n\left[ c_{-n} \b_1(-n) +  
s_{-n} \b_2(n)\right] \Phi\xxe
\nonueqaend
equal to $-\xxa J(\al'')\Phi\xxe$. \QED

\subsection*{C.3 Proof of Lemma~\ref{L3}}
We introduce the notations 
$$
W^3_\pm \, : = \, \lim_{\eps\downarrow 0} \f{1}{6\pi}
\int_{0}^{2\pi} dy\,  \rho^\pm_\eps(y)^3 $$
with $\rho^\pm_\eps$ given in Eq.\ \Ref{bbpm}, and a simple computation yields  
\eqa
\label{Wpm}
W^3_\pm  = \lim_{\eps\downarrow 0}\f{1}{6\pi}\int_{0}^{2\pi} d y 
\sum_{n_1,n_2,n_3=1}^\infty 
\left(c_{\mp n_1}\b_1(\mp n_1)\ee{\mp\ii {n_1}y} +   
s_{\pm n_1} \b_2(\mp n_1)\ee{ \pm \ii {n_1} y}  \right) \nonu \times  
\left( n_1\leftrightarrow n_2 \right)  
\left(  n_2\leftrightarrow n_3  \right)
\ee{-(n_1+n_2+n_3)\eps} = \\
= \sum_{m,n=1}^\infty 
\left( c_{\mp m} c_{\mp n} s_{\pm(m+n)}\b_1(\mp m)\b_1(\mp n)\b_2(\mp m\mp n) 
\right .\nonu \left.  
+  s_{\pm m} s_{\pm n} c_{\mp (m+n)}\b_2(\mp m)\b_2(\mp n)\b_1(\mp m\mp n)  \right) 
\eqaend
where `$(n_1\leftrightarrow n_2)$' means the same term as the previous one but with $n_2$ 
instead of $n_1$. We observe that $\cC\Omega=Q\Omega=\rho^-_\eps(x)\Omega=0$ and thus   
$$
\cH\Omega = \nu W^3_+\Omega\, , \quad (W^3_+)^*= W^3_- 
$$
which implies that \Ref{sq21} is true if and only if
$$
(*)_+ - (*)_-\, := \,   \pr{\Omega}{\Phi W_+^3 \Omega} - 
\pr{\Omega}{ W_-^3 \Phi \Omega}  
$$
vanishes. 
We now compute $(*)_-$ using Eqs.\ \Ref{b121}--\Ref{b122} which imply
(for $n>0$)
$$
\b_1(n) \Phi \Omega = \ii n c_{-n}\alpha_n \Phi \Omega  \, , \quad 
\b_2(n)\Phi \Omega = \ii n s_{n}\alpha_{-n} \Phi \Omega  \: , 
$$
and therefore
\nonueqa
(*)_- = \sum_{m,n=1}^\infty
\bigl<\Omega, [ c_{m} c_{n} s_{-(m+n)} \b_1(m)\b_1(n)\b_2(m+n)
+  \\    
+ s_{-m} s_{-n} c_{m+n} \b_2(m)\b_2(n)\b_1(m+n) ]   
\Phi \Omega\bigr> \, 
= \delta_{w,0} \sum_{m,n=1}^\infty (\ii)^3 mn(m+n) \nonu \times \left( 
|c_m|^2 |c_n|^2 |s_{m+n}|^2 \alpha_m\alpha_n\alpha_{-(m+n)}
+  |s_m|^2 |s_n|^2 |c_{m+n}|^2 \alpha_{-m}\alpha_{-n}\alpha_{m+n}
 \right) 
\nonueqaend
where we used $\pr{\Omega}{\Phi \Omega} \,  = \delta_{w,0}$. 
The equations in \Ref{b122} also imply (for $n>0$)
$$
<\Omega, \Phi\b_1(-n) = \ii n c_{n}\alpha_{-n} <\Omega, \Phi \, , \quad 
<\Omega, \Phi\b_2(-n) = \ii n s_{-n}\alpha_{n} <\Omega, \Phi  \: , 
$$
and thus
\nonueqa
(*)_+ = \sum_{m,n=1}^\infty\bigl<\Omega,\Phi 
[ c_{-m} c_{-n} s_{m+n}\b_1(-m)\b_1(-n)\b_2(-m-n) 
+ \\  
+ s_{m} s_{n} c_{-(m+n)}\b_2(-m)\b_2(-n)\b_1(-m-n)  ]   
 \Omega\bigr> \, = 
 \delta_{w,0} \sum_{m,n=1}^\infty(\ii)^3 mn(m+n)\nonu\times  \left( 
|c_m|^2 |c_n|^2 |s_{m+n}|^2 \alpha_{-m}\alpha_{-n} \alpha_{m+n}
+ |s_m|^2 |s_n|^2 |c_{m+n}|^2 \alpha_{m}\alpha_{n} \alpha_{-(m+n)}
 \right) \: . 
\nonueqaend
We thus obtain
\eqa
(*)_+  -  (*)_- =\delta_{w,0} \sum_{m,n=1}^\infty (\ii)^3 mn(m+n) \nonu \times 
\left( |c_m|^2 |c_n|^2 |s_{m+n}|^2 -  |s_m|^2 |s_n|^2 |c_{m+n}|^2 \right)
\left(
\alpha_{-m}\alpha_{-n} \alpha_{m+n} +  
\alpha_{m}\alpha_{n} \alpha_{-(m+n)} \right)   
\eqaend
which shows explicitly that the relations in Eq.\ \Ref{crucial} imply  
$(*)_+ -(*)_-=0$. \QED

\end{document}